\documentstyle[10pt,epsf,epsfig,dp_delphititle]{dp_delphi}
\makeindex
\def\DpPaperGroup{EP}
\def\DpPaperRef{2000-133}
\def\DpDate{11 July 2000}
\def\DpAuthors{DELPHI Collaboration}
\def\DpSubmit{(Submitted to Eur.Phys.J)}
\def\DpTitle{{
Search for neutralino pair production
at \boldmath $\sqrt{s}$ = 189 GeV
}}
\def\DpComment{ }
\def\DpEMail{ }
\begin{document}
\makeatletter
\newcount\@tempcntc
\def\@citex[#1]#2{\if@filesw\immediate\write\@auxout{\string\citation{#2}}\fi
  \@tempcnta\z@\@tempcntb\m@ne\def\@citea{}\@cite{\@for\@citeb:=#2\do
    {\@ifundefined
       {b@\@citeb}{\@citeo\@tempcntb\m@ne\@citea\def\@citea{,}{\bf ?}\@warning
       {Citation `\@citeb' on page \thepage \space undefined}}%
    {\setbox\z@\hbox{\global\@tempcntc0\csname b@\@citeb\endcsname\relax}%
     \ifnum\@tempcntc=\z@ \@citeo\@tempcntb\m@ne
       \@citea\def\@citea{,}\hbox{\csname b@\@citeb\endcsname}%
     \else
      \advance\@tempcntb\@ne
      \ifnum\@tempcntb=\@tempcntc
      \else\advance\@tempcntb\m@ne\@citeo
      \@tempcnta\@tempcntc\@tempcntb\@tempcntc\fi\fi}}\@citeo}{#1}}
\def\@citeo{\ifnum\@tempcnta>\@tempcntb\else\@citea\def\@citea{,}%
  \ifnum\@tempcnta=\@tempcntb\the\@tempcnta\else
   {\advance\@tempcnta\@ne\ifnum\@tempcnta=\@tempcntb \else \def\@citea{--}\fi
    \advance\@tempcnta\m@ne\the\@tempcnta\@citea\the\@tempcntb}\fi\fi}
 
\makeatother
\begin{titlepage}
\pagenumbering{roman}
\CERNpreprint{\DpPaperGroup}{\DpPaperRef} 
\date{{\small\DpDate}} 
\title{\DpTitle} 
\address{\DpAuthors} 
\begin{shortabs} 
\noindent
A search for pair-production of neutralinos at a LEP
centre-of-mass energy of 189 GeV gave no evidence for
a signal.
This limits the neutralino production cross-section
and excludes regions in the parameter space of the
Minimal Supersymmetric Standard Model (MSSM).
\end{shortabs}
\vfill
\begin{center}
\DpSubmit \ \\ 
\DpComment \ \\
\DpEMail \ \\
\end{center}
\vfill
\clearpage
\headsep 10.0pt
\addtolength{\textheight}{10mm}
\addtolength{\footskip}{-5mm}
\begingroup
%
\newcommand{\DpName}[2]{\hbox{#1$^{\ref{#2}}$},\hfill}
\newcommand{\DpNameTwo}[3]{\hbox{#1$^{\ref{#2},\ref{#3}}$},\hfill}
\newcommand{\DpNameThree}[4]{\hbox{#1$^{\ref{#2},\ref{#3},\ref{#4}}$},\hfill}
\newskip\Bigfill \Bigfill = 0pt plus 1000fill
\newcommand{\DpNameLast}[2]{\hbox{#1$^{\ref{#2}}$}\hspace{\Bigfill}}
%
\footnotesize
\noindent
\DpName{P.Abreu}{LIP}
\DpName{W.Adam}{VIENNA}
\DpName{T.Adye}{RAL}
\DpName{P.Adzic}{DEMOKRITOS}
\DpName{I.Ajinenko}{SERPUKHOV}
\DpName{Z.Albrecht}{KARLSRUHE}
\DpName{T.Alderweireld}{AIM}
\DpName{G.D.Alekseev}{JINR}
\DpName{R.Alemany}{CERN}
\DpName{T.Allmendinger}{KARLSRUHE}
% ...
\normalsize
\endgroup
\titlefoot{Department of Physics and Astronomy, Iowa State
     University, Ames IA 50011-3160, USA
    \label{AMES}}
\titlefoot{Physics Department, Univ. Instelling Antwerpen,
     Universiteitsplein 1, B-2610 Antwerpen, Belgium \\
     \indent~~and IIHE, ULB-VUB,
     Pleinlaan 2, B-1050 Brussels, Belgium \\
     \indent~~and Facult\'e des Sciences,
     Univ. de l'Etat Mons, Av. Maistriau 19, B-7000 Mons, Belgium
    \label{AIM}}
\titlefoot{Physics Laboratory, University of Athens, Solonos Str.
     104, GR-10680 Athens, Greece
    \label{ATHENS}}
\titlefoot{Department of Physics, University of Bergen,
     All\'egaten 55, NO-5007 Bergen, Norway
    \label{BERGEN}}
\titlefoot{Dipartimento di Fisica, Universit\`a di Bologna and INFN,
     Via Irnerio 46, IT-40126 Bologna, Italy
    \label{BOLOGNA}}
\titlefoot{Centro Brasileiro de Pesquisas F\'{\i}sicas, rua Xavier Sigaud 150,
     BR-22290 Rio de Janeiro, Brazil \\
     \indent~~and Depto. de F\'{\i}sica, Pont. Univ. Cat\'olica,
     C.P. 38071 BR-22453 Rio de Janeiro, Brazil \\
     \indent~~and Inst. de F\'{\i}sica, Univ. Estadual do Rio de Janeiro,
     rua S\~{a}o Francisco Xavier 524, Rio de Janeiro, Brazil
    \label{BRASIL}}
\titlefoot{Comenius University, Faculty of Mathematics and Physics,
     Mlynska Dolina, SK-84215 Bratislava, Slovakia
    \label{BRATISLAVA}}
\titlefoot{Coll\`ege de France, Lab. de Physique Corpusculaire, IN2P3-CNRS,
     FR-75231 Paris Cedex 05, France
    \label{CDF}}
\titlefoot{CERN, CH-1211 Geneva 23, Switzerland
    \label{CERN}}
\titlefoot{Institut de Recherches Subatomiques, IN2P3 - CNRS/ULP - BP20,
     FR-67037 Strasbourg Cedex, France
    \label{CRN}}
\titlefoot{Now at DESY-Zeuthen, Platanenallee 6, D-15735 Zeuthen, Germany
    \label{DESY}}
\titlefoot{Institute of Nuclear Physics, N.C.S.R. Demokritos,
     P.O. Box 60228, GR-15310 Athens, Greece
    \label{DEMOKRITOS}}
\titlefoot{FZU, Inst. of Phys. of the C.A.S. High Energy Physics Division,
     Na Slovance 2, CZ-180 40, Praha 8, Czech Republic
    \label{FZU}}
\titlefoot{Dipartimento di Fisica, Universit\`a di Genova and INFN,
     Via Dodecaneso 33, IT-16146 Genova, Italy
    \label{GENOVA}}
\titlefoot{Institut des Sciences Nucl\'eaires, IN2P3-CNRS, Universit\'e
     de Grenoble 1, FR-38026 Grenoble Cedex, France
    \label{GRENOBLE}}
\titlefoot{Helsinki Institute of Physics, HIP,
     P.O. Box 9, FI-00014 Helsinki, Finland
    \label{HELSINKI}}
\titlefoot{Joint Institute for Nuclear Research, Dubna, Head Post
     Office, P.O. Box 79, RU-101 000 Moscow, Russian Federation
    \label{JINR}}
\titlefoot{Institut f\"ur Experimentelle Kernphysik,
     Universit\"at Karlsruhe, Postfach 6980, DE-76128 Karlsruhe,
     Germany
    \label{KARLSRUHE}}
\titlefoot{Institute of Nuclear Physics and University of Mining and Metalurgy,
     Ul. Kawiory 26a, PL-30055 Krakow, Poland
    \label{KRAKOW}}
\titlefoot{Universit\'e de Paris-Sud, Lab. de l'Acc\'el\'erateur
     Lin\'eaire, IN2P3-CNRS, B\^{a}t. 200, FR-91405 Orsay Cedex, France
    \label{LAL}}
\titlefoot{School of Physics and Chemistry, University of Lancaster,
     Lancaster LA1 4YB, UK
    \label{LANCASTER}}
\titlefoot{LIP, IST, FCUL - Av. Elias Garcia, 14-$1^{o}$,
     PT-1000 Lisboa Codex, Portugal
    \label{LIP}}
\titlefoot{Department of Physics, University of Liverpool, P.O.
     Box 147, Liverpool L69 3BX, UK
    \label{LIVERPOOL}}
\titlefoot{LPNHE, IN2P3-CNRS, Univ.~Paris VI et VII, Tour 33 (RdC),
     4 place Jussieu, FR-75252 Paris Cedex 05, France
    \label{LPNHE}}
\titlefoot{Department of Physics, University of Lund,
     S\"olvegatan 14, SE-223 63 Lund, Sweden
    \label{LUND}}
\titlefoot{Universit\'e Claude Bernard de Lyon, IPNL, IN2P3-CNRS,
     FR-69622 Villeurbanne Cedex, France
    \label{LYON}}
\titlefoot{Univ. d'Aix - Marseille II - CPP, IN2P3-CNRS,
     FR-13288 Marseille Cedex 09, France
    \label{MARSEILLE}}
\titlefoot{Dipartimento di Fisica, Universit\`a di Milano and INFN-MILANO,
     Via Celoria 16, IT-20133 Milan, Italy
    \label{MILANO}}
\titlefoot{Dipartimento di Fisica, Univ. di Milano-Bicocca and
     INFN-MILANO, Piazza delle Scienze 2, IT-20126 Milan, Italy
    \label{MILANO2}}
\titlefoot{IPNP of MFF, Charles Univ., Areal MFF,
     V Holesovickach 2, CZ-180 00, Praha 8, Czech Republic
    \label{NC}}
\titlefoot{NIKHEF, Postbus 41882, NL-1009 DB
     Amsterdam, The Netherlands
    \label{NIKHEF}}
\titlefoot{National Technical University, Physics Department,
     Zografou Campus, GR-15773 Athens, Greece
    \label{NTU-ATHENS}}
\titlefoot{Physics Department, University of Oslo, Blindern,
     NO-1000 Oslo 3, Norway
    \label{OSLO}}
\titlefoot{Dpto. Fisica, Univ. Oviedo, Avda. Calvo Sotelo
     s/n, ES-33007 Oviedo, Spain
    \label{OVIEDO}}
\titlefoot{Department of Physics, University of Oxford,
     Keble Road, Oxford OX1 3RH, UK
    \label{OXFORD}}
\titlefoot{Dipartimento di Fisica, Universit\`a di Padova and
     INFN, Via Marzolo 8, IT-35131 Padua, Italy
    \label{PADOVA}}
\titlefoot{Rutherford Appleton Laboratory, Chilton, Didcot
     OX11 OQX, UK
    \label{RAL}}
\titlefoot{Dipartimento di Fisica, Universit\`a di Roma II and
     INFN, Tor Vergata, IT-00173 Rome, Italy
    \label{ROMA2}}
\titlefoot{Dipartimento di Fisica, Universit\`a di Roma III and
     INFN, Via della Vasca Navale 84, IT-00146 Rome, Italy
    \label{ROMA3}}
\titlefoot{DAPNIA/Service de Physique des Particules,
     CEA-Saclay, FR-91191 Gif-sur-Yvette Cedex, France
    \label{SACLAY}}
\titlefoot{Instituto de Fisica de Cantabria (CSIC-UC), Avda.
     los Castros s/n, ES-39006 Santander, Spain
    \label{SANTANDER}}
\titlefoot{Dipartimento di Fisica, Universit\`a degli Studi di Roma
     La Sapienza, Piazzale Aldo Moro 2, IT-00185 Rome, Italy
    \label{SAPIENZA}}
\titlefoot{Inst. for High Energy Physics, Serpukov
     P.O. Box 35, Protvino, (Moscow Region), Russian Federation
    \label{SERPUKHOV}}
\titlefoot{J. Stefan Institute, Jamova 39, SI-1000 Ljubljana, Slovenia
     and Laboratory for Astroparticle Physics,\\
     \indent~~Nova Gorica Polytechnic, Kostanjeviska 16a, SI-5000 Nova Gorica, Slovenia, \\
     \indent~~and Department of Physics, University of Ljubljana,
     SI-1000 Ljubljana, Slovenia
    \label{SLOVENIJA}}
\titlefoot{Fysikum, Stockholm University,
     Box 6730, SE-113 85 Stockholm, Sweden
    \label{STOCKHOLM}}
\titlefoot{Dipartimento di Fisica Sperimentale, Universit\`a di
     Torino and INFN, Via P. Giuria 1, IT-10125 Turin, Italy
    \label{TORINO}}
\titlefoot{Dipartimento di Fisica, Universit\`a di Trieste and
     INFN, Via A. Valerio 2, IT-34127 Trieste, Italy \\
     \indent~~and Istituto di Fisica, Universit\`a di Udine,
     IT-33100 Udine, Italy
    \label{TU}}
\titlefoot{Univ. Federal do Rio de Janeiro, C.P. 68528
     Cidade Univ., Ilha do Fund\~ao
     BR-21945-970 Rio de Janeiro, Brazil
    \label{UFRJ}}
\titlefoot{Department of Radiation Sciences, University of
     Uppsala, P.O. Box 535, SE-751 21 Uppsala, Sweden
    \label{UPPSALA}}
\titlefoot{IFIC, Valencia-CSIC, and D.F.A.M.N., U. de Valencia,
     Avda. Dr. Moliner 50, ES-46100 Burjassot (Valencia), Spain
    \label{VALENCIA}}
\titlefoot{Institut f\"ur Hochenergiephysik, \"Osterr. Akad.
     d. Wissensch., Nikolsdorfergasse 18, AT-1050 Vienna, Austria
    \label{VIENNA}}
\titlefoot{Inst. Nuclear Studies and University of Warsaw, Ul.
     Hoza 69, PL-00681 Warsaw, Poland
    \label{WARSZAWA}}
\titlefoot{Fachbereich Physik, University of Wuppertal, Postfach
     100 127, DE-42097 Wuppertal, Germany
    \label{WUPPERTAL}}
\addtolength{\textheight}{-10mm}
\addtolength{\footskip}{5mm}
\clearpage
\headsep 30.0pt
\end{titlepage}
%
\pagenumbering{arabic} 
\setcounter{footnote}{0} %
\large

%

\def\leqsim{\mathbin{\;\raise1pt\hbox{$<$}\kern-8pt\lower3pt\hbox{\small$\sim$}\;}}
\def\geqsim{\mathbin{\;\raise1pt\hbox{$>$}\kern-8pt\lower3pt\hbox{\small$\sim$}\;}}
\newcommand{\dfrac}[2]{\frac{\displaystyle #1}{\displaystyle #2}}
\renewcommand\topfraction{1.}
\renewcommand\bottomfraction{1.}
\renewcommand\floatpagefraction{0.}
\renewcommand\textfraction{0.}
\def\MXN#1{\mbox{$ M_{\tilde{\chi}^0_#1}                                $}}
\def\MXNN#1#2{\mbox{$ M_{\tilde{\chi}^0_{#1,#2}}                        $}}
\def\MXNNN#1#2#3{\mbox{$ M_{\tilde{\chi}^0_{#1,#2,#3}}                  $}}
\def\MXC#1{\mbox{$ M_{\tilde{\chi}^{\pm}_#1}                            $}}
\def\XP#1{\mbox{$ \tilde{\chi}^+_#1                                     $}}
\def\XPP#1#2{\mbox{$ \tilde{\chi}^{+}_{#1,#2}                           $}}
\def\XCC#1#2{\mbox{$ \tilde{\chi}^{-}_{#1,#2}                             $}}
\def\XM#1{\mbox{$ \tilde{\chi}^-_#1                                     $}}
\def\XPM#1{\mbox{$ \tilde{\chi}^{\pm}_#1                                $}}
\def\XN#1{\mbox{$ \tilde{\chi}^0_#1                                     $}}
\def\XNN#1#2{\mbox{$ \tilde{\chi}^0_{#1,#2}                             $}}
\def\XNNN#1#2#3{\mbox{$ \tilde{\chi}^0_{#1,#2,#3}                       $}}
\def\p#1{\mbox{$ \mbox{\bf p}_1                                         $}}
\newcommand{\Gino}    {\mbox{$ \tilde{\mathrm G}                           $}}
\newcommand{\tanb}    {\mbox{$ \tan \beta                                  $}}
\newcommand{\smu}     {\mbox{$ \tilde{\mu}                                 $}}
\newcommand{\msmu}    {\mbox{$ M_{\tilde{\mu}}                             $}}
\newcommand{\msmur}   {\mbox{$ M_{\tilde{\mu}_R}                           $}}
\newcommand{\msmul}   {\mbox{$ M_{\tilde{\mu}_L}                           $}}
\newcommand{\sel}     {\mbox{$ \tilde{\mathrm e}                           $}}
\newcommand{\msel}    {\mbox{$ M_{\tilde{\mathrm e}}                       $}}
\newcommand{\stau}    {\mbox{$ \tilde{\tau}                                $}}
\newcommand{\stone}   {\mbox{$ \tilde{\tau}_1                              $}}
\newcommand{\sttwo}   {\mbox{$ \tilde{\tau}_2                              $}}
\newcommand{\mstau}   {\mbox{$ M_{\tilde{\tau}}                            $}}
\newcommand{\mstone}  {\mbox{$ M_{\tilde{\tau}_1}                          $}}
\newcommand{\msttwo}  {\mbox{$ M_{\tilde{\tau}_2}                          $}}
\newcommand{\snu}     {\mbox{$ \tilde\nu                                   $}}
\newcommand{\msnu}    {\mbox{$ M_{\tilde\nu}                               $}}
\newcommand{\msell}   {\mbox{$ M_{\tilde{\mathrm e}_L}                     $}}
\newcommand{\mselr}   {\mbox{$ M_{\tilde{\mathrm e}_R}                     $}}
\newcommand{\sfe}     {\mbox{$ \tilde{\mathrm f}                           $}}
\newcommand{\msfe}    {\mbox{$ M_{\tilde{\mathrm f}}                       $}}
\newcommand{\sle}     {\mbox{$ \tilde{\ell}                                $}}
\newcommand{\msle}    {\mbox{$ M_{\tilde{\ell}}                            $}}
\newcommand{\stq}     {\mbox{$ \tilde {\mathrm t}                          $}}
\newcommand{\mstq}    {\mbox{$ M_{\tilde {\mathrm t}}                      $}}
\newcommand{\sbq}     {\mbox{$ \tilde {\mathrm b}                          $}}
\newcommand{\msbq}    {\mbox{$ M_{\tilde {\mathrm b}}                      $}}
\newcommand{\An}      {\mbox{$ {\mathrm A}^0                               $}}
\newcommand{\hn}      {\mbox{$ {\mathrm h}^0                               $}}
\newcommand{\Zn}      {\mbox{$ {\mathrm Z}                                 $}}
\newcommand{\Zstar}   {\mbox{$ {\mathrm Z}^*                               $}}
\newcommand{\Hn}      {\mbox{$ {\mathrm H}^0                               $}}
\newcommand{\HP}      {\mbox{$ {\mathrm H}^+                               $}}
\newcommand{\HM}      {\mbox{$ {\mathrm H}^-                               $}}
\newcommand{\Wp}      {\mbox{$ {\mathrm W}^+                               $}}
\newcommand{\Wm}      {\mbox{$ {\mathrm W}^-                               $}}
\newcommand{\Wstar}   {\mbox{$ {\mathrm W}^*                               $}}
\newcommand{\WW}      {\mbox{$ {\mathrm W}^+{\mathrm W}^-                  $}}
\newcommand{\ZZ}      {\mbox{$ {\mathrm Z}{\mathrm Z}                      $}}
\newcommand{\HZ}      {\mbox{$ {\mathrm H}^0 {\mathrm Z}                   $}}
\newcommand{\GW}      {\mbox{$ \Gamma_{\mathrm W}                          $}}
\newcommand{\Zg}      {\mbox{$ \Zn \gamma                                  $}}
\newcommand{\sqs}     {\mbox{$ \sqrt{s}                                    $}}
\newcommand{\epm}     {\mbox{$ {\mathrm e}^{\pm}                           $}}
\newcommand{\ee}      {\mbox{$ {\mathrm e}^+ {\mathrm e}^-                 $}}
\newcommand{\mumu}    {\mbox{$ \mu^+ \mu^-                                 $}}
\newcommand{\tautau}  {\mbox{$ \tau^+ \tau^-                               $}}
\newcommand{\eeto}    {\mbox{$ {\mathrm e}^+ {\mathrm e}^-\! \to\          $}}
\newcommand{\ellell}  {\mbox{$ \ell^+ \ell^-                               $}}
\newcommand{\eeWW}    {\mbox{$ \ee \rightarrow \WW                         $}}
\newcommand{\eV}      {\mbox{$ {\mathrm{eV}}                               $}}
\newcommand{\eVc}     {\mbox{$ {\mathrm{eV}}/c                             $}}
\newcommand{\eVcc}    {\mbox{$ {\mathrm{eV}}/c^2                           $}}
\newcommand{\MeV}     {\mbox{$ {\mathrm{MeV}}                              $}}
\newcommand{\MeVc}    {\mbox{$ {\mathrm{MeV}}/c                            $}}
\newcommand{\MeVcc}   {\mbox{$ {\mathrm{MeV}}/c^2                          $}}
\newcommand{\GeV}     {\mbox{$ {\mathrm{GeV}}                              $}}
\newcommand{\GeVc}    {\mbox{$ {\mathrm{GeV}}/c                            $}}
\newcommand{\GeVcc}   {\mbox{$ {\mathrm{GeV}}/c^2                          $}}
\newcommand{\TeV}     {\mbox{$ {\mathrm{TeV}}                              $}}
\newcommand{\TeVc}    {\mbox{$ {\mathrm{TeV}}/c                            $}}
\newcommand{\TeVcc}   {\mbox{$ {\mathrm{TeV}}/c^2                          $}}
\newcommand{\pbi}     {\mbox{$ {\mathrm{pb}}^{-1}                          $}}
\newcommand{\MZ}      {\mbox{$ M_{\mathrm Z}                               $}}
\newcommand{\MW}      {\mbox{$ M_{\mathrm W}                               $}}
\newcommand{\MA}      {\mbox{$ m_{\mathrm A}                               $}}
\newcommand{\GF}      {\mbox{$ {\mathrm G}_{\mathrm F}                     $}}
\newcommand{\MH}      {\mbox{$ m_{{\mathrm H}^0}                           $}}
\newcommand{\MHP}     {\mbox{$ m_{{\mathrm H}^\pm}                         $}}
\newcommand{\MSH}     {\mbox{$ m_{{\mathrm h}^0}                           $}}
\newcommand{\MT}      {\mbox{$ m_{\mathrm t}                               $}}
\newcommand{\GZ}      {\mbox{$ \Gamma_{{\mathrm Z} }                       $}}
\newcommand{\SS}      {\mbox{$ \mathrm S                                   $}}
\newcommand{\TT}      {\mbox{$ \mathrm T                                   $}}
\newcommand{\UU}      {\mbox{$ \mathrm U                                   $}}
\newcommand{\alphmz}  {\mbox{$ \alpha (m_{{\mathrm Z}})                    $}}
\newcommand{\alphas}  {\mbox{$ \alpha_{\mathrm s}                          $}}
\newcommand{\alphmsb} {\mbox{$ \alphas (m_{\mathrm Z})
                               _{\overline{\mathrm{MS}}}                   $}}
\newcommand{\alphbar} {\mbox{$ \overline{\alpha}_{\mathrm s}               $}}
\newcommand{\Ptau}    {\mbox{$ P_{\tau}                                    $}}
\newcommand{\mean}[1] {\mbox{$ \left\langle #1 \right\rangle               $}}
\newcommand{\dgree}   {\mbox{$ ^\circ                                      $}}
\newcommand{\qqg}     {\mbox{$ {\mathrm q}\bar{\mathrm q}\gamma            $}}
\newcommand{\Wev}     {\mbox{$ {\mathrm{W e}} \nu_{\mathrm e}              $}}
\newcommand{\Zvv}     {\mbox{$ \Zn \nu \bar{\nu}                           $}}
\newcommand{\Zee}     {\mbox{$ \Zn \ee                                     $}}
\newcommand{\ctw}     {\mbox{$ \cos\theta_{\mathrm W}                      $}}
\newcommand{\thw}     {\mbox{$ \theta_{\mathrm W}                          $}}
\newcommand{\thetabar}{\mbox{$ \theta^*                                    $}}
\newcommand{\phibar}  {\mbox{$ \phi^*                                      $}}
\newcommand{\thetapl} {\mbox{$ \theta_+                                    $}}
\newcommand{\phipl}   {\mbox{$ \phi_+                                      $}}
\newcommand{\thetamin}{\mbox{$ \theta_-                                    $}}
\newcommand{\phimin}  {\mbox{$ \phi_-                                      $}}
\newcommand{\ds}      {\mbox{$ {\mathrm d} \sigma                          $}}
\def    \ll           {\mbox{$\ell \ell                                    $}}
\def    \jjl          {\mbox{$jj \ell                           $}}
\def    \jj           {\mbox{$jj                                $}}
\def   \jjjj          {\mbox{${\it jets}                                   $}}
\newcommand{\jjlv}    {\mbox{$ j j \ell \nu                                $}}
\newcommand{\jjvv}    {\mbox{$ j j \nu \bar{\nu}                           $}}
\newcommand{\qqvv}    {\mbox{$ \mathrm{q \bar{q}} \nu \bar{\nu}            $}}
\newcommand{\qqll}    {\mbox{$ \mathrm{q \bar{q}} \ell \bar{\ell}          $}}
\newcommand{\jjll}    {\mbox{$ j j \ell \bar{\ell}                         $}}
\newcommand{\lvlv}    {\mbox{$ \ell \nu \ell \nu                           $}}
\newcommand{\dz}      {\mbox{$ \delta g_{\mathrm{W W Z}    }               $}}
\newcommand{\pT}      {\mbox{$ p_{\mathrm{T}}                              $}}
\newcommand{\pL}      {\mbox{$ p_{\mathrm{L}}                              $}}
\newcommand{\thetap}  {\mbox{$ \theta_{p}                                  $}}
\newcommand{\ptr}     {\mbox{$ p_{\perp}                                   $}}
\newcommand{\ptrjet}  {\mbox{$ p_{\perp {\mathrm{jet}}}                    $}}
\newcommand{\Wvis}    {\mbox{$ {\mathrm W}_{\mathrm{vis}}                  $}}
\newcommand{\gamgam}  {\mbox{$ \gamma \gamma                               $}}
\newcommand{\qaqb}    {\mbox{$ {\mathrm q}_1 \bar{\mathrm q}_2             $}}
\newcommand{\qcqd}    {\mbox{$ {\mathrm q}_3 \bar{\mathrm q}_4             $}}
\newcommand{\bbbar}   {\mbox{$ {\mathrm b}\bar{\mathrm b}                  $}}
\newcommand{\ffbar}   {\mbox{$ {\mathrm f}\bar{\mathrm f}                  $}}
\newcommand{\ffbarp}  {\mbox{$ {\mathrm f}\bar{\mathrm f}'                 $}}
\newcommand{\fpfbarp} {\mbox{$ {\mathrm f}'\bar{\mathrm f}'                $}}
\newcommand{\qqbar}   {\mbox{$ {\mathrm q}\bar{\mathrm q}                  $}}
\newcommand{\nunubar} {\mbox{$ {\nu}\bar{\nu}                              $}}
\newcommand{\qqbarp}  {\mbox{$ {\mathrm q'}\bar{\mathrm q}'                $}}
\newcommand{\djoin}   {\mbox{$ d_{\mathrm{join}}                           $}}
\newcommand{\mErad}   {\mbox{$ \left\langle E_{\mathrm{rad}} \right\rangle $}}
\newcommand{\Lum}{${\cal L}\;$}
\newcommand{\lum}{{\cal L}}
\newcommand{\Cms}{$\mbox{ cm}^{-2} \mbox{ s}^{-1}\;$}
\newcommand{\cms}{\mbox{ cm}^{-2} \mbox{ s}^{-1}\;}
\newcommand{\Ecms}    {\mbox{$ E_{\mathrm{\small cms}}                      $}}
\newcommand{\Evis}    {\mbox{$ E_{\mathrm{\small vis}}                      $}}
\newcommand{\ET}      {\mbox{$ E_{\mathrm{\small T}}                        $}}
\newcommand{\Erec}    {\mbox{$ E_{\mathrm{\small rec}}                      $}}
\newcommand{\Erad}    {\mbox{$ E_{\mathrm{\small rad}}                      $}}
\newcommand{\Mvis}    {\mbox{$ M_{\mathrm{\small vis}}                      $}}
\newcommand{\Mrec}    {\mbox{$ M_{\mathrm{\small rec}}                      $}}
\newcommand{\pvis}    {\mbox{$ p_{\mathrm{\small vis}}                      $}}
\newcommand{\Minv}    {\mbox{$ M_{\mathrm{\small inv}}                      $}}
\newcommand{\pmiss}   {\mbox{$ p_{\mathrm{\small miss}}                     $}}
\newcommand{\ptmiss}  {\mbox{$ p_T^{\mathrm{\small miss}}                   $}}
\newcommand{\ptpair}  {\mbox{$ p_T^{\mathrm{\small pair}}                   $}}
\newcommand{\Mhfit}{\; \hat{m}_{H^0} }
\newcommand{\bl}      {\mbox{\ \ \ \ \ \ \ \ \ \ } }
\newcommand{\Zto}   {\mbox{$\mathrm Z^0 \to$}}
\newcommand{\etal}  {\mbox{\it et al.}}
\def\NPB#1#2#3{{\rm Nucl.~Phys.} {\bf{B#1}} (19#2) #3}
\def\PLB#1#2#3{{\rm Phys.~Lett.} {\bf{B#1}} (#2) #3}
\def\PRD#1#2#3{{\rm Phys.~Rev.} {\bf{D#1}} (19#2) #3}
\def\PRL#1#2#3{{\rm Phys.~Rev.~Lett.} {\bf{#1}} (19#2) #3}
\def\ZPC#1#2#3{{\rm Z.~Phys.} {\bf C#1} (19#2) #3}
\def\EPJC#1#2#3{{\rm E.~Phys.~J.} {\bf C#1} (19#2) #3}
\def\PTP#1#2#3{{\rm Prog.~Theor.~Phys.} {\bf#1}  (19#2) #3}
\def\MPL#1#2#3{{\rm Mod.~Phys.~Lett.} {\bf#1} (19#2) #3}
\def\PR#1#2#3{{\rm Phys.~Rep.} {\bf#1} (19#2) #3}
\def\RMP#1#2#3{{\rm Rev.~Mod.~Phys.} {\bf#1} (19#2) #3}
\def\HPA#1#2#3{{\rm Helv.~Phys.~Acta} {\bf#1} (19#2) #3}
\def\NIMA#1#2#3{{\rm Nucl.~Instr.~and~Meth.} {\bf A#1} (19#2) #3}
\def\CPC#1#2#3{{\rm Comp.~Phys.~Comm.} {\bf#1} (19#2) #3}
\def    \DM          {\mbox{$\Delta M$}}
\def    \missEt      {\ifmmode{/\mkern-11mu E_t}\else{${/\mkern-11mu E_t}$}\fi}
\def    \missE       {\ifmmode{/\mkern-11mu E}\else{${/\mkern-11mu E}$}\fi}
\def    \missp       {\ifmmode{/\mkern-11mu p}\else{${/\mkern-11mu p}$}\fi}
\def    \misspt      {\ifmmode{/\mkern-11mu p_t}\else{${/\mkern-11mu p_t}$}\fi}
\def    \DML         {\mbox{5~GeV $<\Delta M<$ 10~GeV}}
\def    \rs          {\mbox{$\sqrt{s}$}}
\def    \msneu       {\mbox{$m_{\tilde{\nu}}$}}

\section{Introduction \label{sec:INTRO}}

During 1998, the DELPHI experiment at LEP  accumulated an integrated
luminosity of 158~\pbi\ at a centre-of-mass energy, \sqs, of 188.7~GeV.
Results
of a search for neutralino pair production in these data are reported here.
In a separate
letter \cite{LSPLIM},
these results are interpreted together
with those of other DELPHI searches to set mass limits on neutralinos, sleptons and
charginos.

In the Minimal Supersymmetric Standard Model (the MSSM) \cite{SUSY}, there are four
neutralinos $\XN{i},i=1,4$, numbered in order of increasing mass, and two
charginos $\XPM{j},j=1,2$. These are linear combinations
of the supersymmetric (SUSY) partners of
neutral and charged gauge and Higgs bosons. In the following, $R$-parity conservation
is assumed, implying a stable lightest supersymmetric particle (LSP) which is
assumed to be the \XN{1}. $R$-parity conservation also implies 
pair-production of
SUSY particles, each decaying (directly or indirectly) into a \XN{1},
which is weakly interacting and escapes detection, giving a signature of missing 
energy and momentum.

The neutralinos can be pair-produced at LEP2 via $s$-channel \Zn\ exchange
or $t$-channel exchange of a scalar electron (selectron, \sel).
The decay of heavier
neutralino states to lighter ones typically involves emission of either a 
fermion-antifermion (\ffbar) 
pair or a photon. If the scalar leptons (sleptons) are light, 
the two-body decay $\XN{i} \to \sle \overline{\ell}$
(followed by $\sle \to \XN{j}\ell$) may dominate. Decays via charginos are also possible.

Of the detectable pair production
channels ({\em i.e.} excluding \XN{1}\XN{1}), \XN{1}\XN{2} and \XN{1}\XN{3} are
important for large regions in the parameter space. For a more complete
coverage, however, one must also consider channels
like \XN{2}\XN{3} and \XN{2}\XN{4}, giving cascade decays with multiple jets or
leptons in the final state.

Moreover, a light scalar tau
lepton (stau, \stau) is likely to arise because of left-right mixing of the stau
states. If the mass of the lighter stau, \mstone, is close to \MXN{1}, the decay
of the \stone\ gives an undetectable neutralino and a low energy $\tau$ which is 
difficult to detect. In this case the search for chargino pair-production 
has a low efficiency since \XPM{1} decays into \stone$\nu$, but the \XN{1}\XN{2} 
and \XN{2}\XN{2} channels are still detectable because of the $\tau$ 
produced directly in the decay $\XN{2}\! \to\! \stone\tau$. It is therefore 
important to search also for these channels \cite{LEPSUSYWG}.

In the search for $\XN{k} \XN{1}$ production with $\XN{k} \to \XN{1} + \ffbar$,
the methods described in Refs. \cite{charneut183,CHANEUT} were used, with minor
changes. 
The signatures consist of pairs of jets or leptons with high missing
energy and momentum and large acoplanarity \footnote{This is defined as the
the complement with respect to 180\dgree\ of the angle between the jet- 
or lepton momenta projected on a plane transverse to the beam axis.}.
In addition, several new searches were introduced in order to obtain a more
complete coverage, in particular in the regions of low \MXN{1}:

\begin{itemize}
\item A search for multijet events, for example from $\XN{i}\XN{j}$$(i=1,2,j=3,4)$
      with \XN{j}~$\to$~\XN{2}\qqbar\ and \XN{2}~ decaying to \XN{1}\qqbar\ or
      \XN{1}$\gamma$.
\item A search for multilepton events for the corresponding decays to lepton pairs.
\item A search for cascade decays with tau leptons, e.g.
      \XN{2}\XN{1} production with \XN{2}~$\to$~\stau$\tau$ and
\stau~$\to$~\XN{1}$\tau$
\item A search for events with low transverse energy and low multiplicity, e.g.
      arising from \XN{2}\XN{2} production with \XN{2}~$\to$~\XN{1}\ellell
      and low $\MXN{2}\! -\! \MXN{1}$,
      or from neutralino decays via intermediate slepton states.
\end{itemize}

The results 
showed no indication of a signal and were used to 
derive limits
within the MSSM scheme with universal parameters at the
high mass scale typical of Grand Unified Theories \cite{SUSY}.

The DELPHI detector has been described elsewhere \cite{DELPHI}.
The central tracking system consists of
a Time Projection Chamber (TPC) and a system of silicon
tracking detectors and drift chambers. 
The electromagnetic
calorimeters are symmetric around the plane perpendicular to the
beam ($\theta$=90\dgree), with the High density Projection Chamber (HPC)
in the barrel region ($\theta\! >\! 43\dgree$) and the 
Forward Electromagnetic Calorimeter (FEMC)
overlapping with the Small angle Tile Calorimeter (STIC) in
the forward region ($1.7\dgree\! <\! \theta\! <\! 35\dgree$). 
The region of poor
electromagnetic calorimetry at a polar angle close to 40\dgree\ is
instrumented by scintillators (hermeticity taggers) which serve to reject
events with unmeasured photons.

\section{Data samples and event generators \label{sec:SAMPLES}}

The total integrated luminosity collected by DELPHI during 1998 at 
\sqs~=~188.7~GeV was
158~\pbi, with 153~\pbi of adequate data quality to be used in the present searches.

To evaluate the signal efficiencies and background contaminations, events were
generated using several different programs. All relied on {\tt JETSET}
7.4~\cite{JETSET}, tuned to LEP~1 data ~\cite{TUNE}, for quark fragmentation.

{\tt SUSYGEN 2.2004} \cite{SUSYGEN} was used to generate neutralino signal events and
calculate cross-sections and branching ratios.

The background process \eeto\qqbar ($n\gamma$) was generated with
{\tt PYTHIA 5.7} \cite{JETSET}.
For $\mu^+\mu^-(\gamma)$ and $\tau^+\tau^-(\gamma)$,
{\tt DYMU3}~\cite{DYMU3} and {\tt KORALZ 4.2}~\cite{KORALZ} were used,
respectively, while the generator of Ref.~\cite{BAFO} was used for \eeto\ee\ events.
Four-fermion final states were generated using 
{\tt EXCALIBUR}~\cite{EXCALIBUR} and {\tt grc4f}~\cite{GRC4F}.

Two-photon interactions giving hadronic final states were generated using
{\tt TWOGAM}~\cite{TWOGAM}, and {\tt PHOJET}~\cite{PHOJET}, while
for those giving leptonic final states the generator of
Ref.~\cite{BDK} was used, including radiative corrections for \ee \mumu\ and
\ee \tautau\ final states.

The generated signal and background events were passed through the
detailed simulation of the DELPHI detector~\cite{DELPHI} and then processed
with the same reconstruction and analysis programs as the real data.
The numbers of simulated events from different background processes were
several times the numbers in the real data, except for the number of
simulated \ee\ pairs from two-photon interactions which was only slightly 
larger than the number expected in the data. 

In addition the simplified fast simulation program {\tt SGV}, previously
used in Ref.~\cite{SGV}, was adopted. {\tt SGV} takes
into account inefficiencies and measurement errors in the
different tracking detectors
and calorimeters, as well as multiple scattering and the showering of
electrons and photons in the tracking volume. This made it possible to
estimate efficiencies for points in the MSSM parameter space
without full simulation, and to take
into account all contributing production and decay channels for a given
point. 

\section{Event selection \label{sec:SELECT}}

The criteria for event selection described below
were based on comparisons of simulated
signal and background event samples.
The different searches used were designed to be mutually exclusive, in order
to allow easy combination of the results.
All searches used the information from the hermeticity taggers to
reject events with photons from initial state radiation lost in the
otherwise insensitive region at polar angles around 40\dgree\ and 140\dgree.
Events were rejected if there were active taggers in the direction of the
missing momentum and not associated to reconstructed jets.
Jets were reconstructed using the LUCLUS algorithm \cite{JETSET} with
$d_{\mathrm join}\!~=\!~10~\GeVc$.
Leptons were identified using the standard DELPHI ``loose tag''
criteria \cite{DELPHI}, except for electrons in the acoplanar leptons
search (see section \ref{sub:2L}).
There is a set of global event variables common to several searches.
These were calculated based on the well-reconstructed particles in
the event and include the total visible energy (\Evis), the
visible mass (\Mvis), total momentum transverse and longitudinal 
to the beam (\pT,\pL), and transverse energy (\ET). The latter
is defined as $\Sigma E_i \sin \theta_i$, where $E_i$ and $\theta_i$
are the energy and polar angle of particle $i$. 
In several cases with two jets, their scaled acoplanarity
(the acoplanarity multiplied by the
sine of the smallest angle between a jet and the beam axis) 
was used.

\subsection{Acoplanar jets search\label{sub:2J}}

Earlier variations of this search at lower energies have been described
in Refs.~\cite{charneut183,CHANEUT}.

At least five well reconstructed charged particles were required, including
at least one with a transverse
momentum with respect to the beam above 1.5~\GeVc. The sum of the 
moduli of momenta of well reconstructed charged
particles had to be greater than 4~\GeVc, and the total transverse energy 
had to exceed 4~GeV.
Two jets were required, each satisfying
$10\dgree\!~<\!~\theta_{\mathrm \small jet}\!~<\!~170\dgree$
and containing at least one well reconstructed charged particle. Tracks
which were badly reconstructed, or did not originate from the 
interaction point, were required not to carry more energy than 0.45\Evis, 
where \Evis\ is the visible energy of well-reconstructed particles. 
This requirement typically removes events with a single badly reconstructed
track with a very high momentum.
In addition, the calorimeter energy associated to such tracks 
had to be less than 0.2\Evis\ for an event to be accepted.

Several criteria were used to reject two-photon events: the fraction
of the total energy carried  by particles emitted within 30\dgree\ of the 
beam had to
be less than 60\%, the polar angle of the total momentum had to satisfy
$|\cos \thetap| < 0.9$, and its transverse component had to exceed 6~\GeVc.

\begin{figure}[ht]
\begin{center}
\mbox{\epsfysize=18.0cm\epsffile{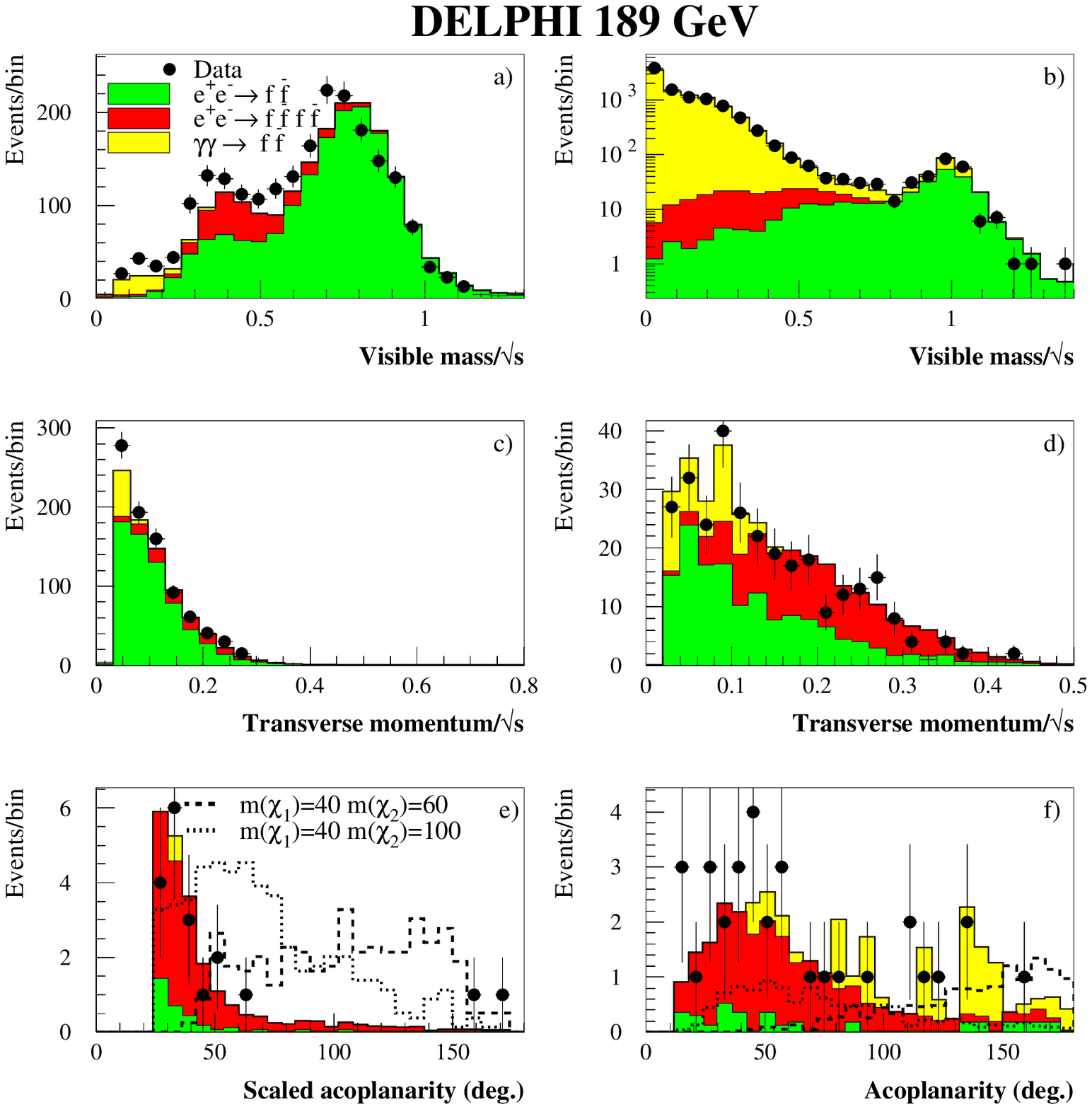}}
\caption[Data-MC comparison]{
The comparison between the real and simulated data
for the acoplanar jet selection (a,c,e) and acoplanar lepton selection
(b,d,f) is shown.
 Plots (a,b) show the visible mass divided
by the centre-of-mass energy at an initial stage of the selections.
Plots (c,e) shows the missing transverse momentum
divided by centre-of-mass energy at an intermediate stage of the
selections.
Plots (e,f) show acoplanarity distributions
after the last step of the selections.
The selections are described in sections \ref{sub:2J} and
\ref{sub:2L}.
Plots (e,f) also show the
expected signal of $\XN{1} \XN{2}$ production
for two different neutralino mass combinations assuming
a cross-section of 1 pb and
the decay $\XN{2} \to Z^* \XN{1}$.

 }
\label{fig:DATAMC1}
\end{center}
\end{figure}

Figure \ref{fig:DATAMC1}(a) shows the distributions of
invariant mass of the visible system (\Mvis) divided by $\sqrt{s}$,
for real and simulated events passing the above selection. 
Here,
and in the following, the simulated sample has been normalised 
to the integrated luminosity used for the data. 
As can be seen from the figure there is some 
excess of data events in the energy region 
corresponding to on-shell Z production with a lost photon from
initial state radiation (``radiative return events''). This can be
ascribed partly to a 3\% deficit in the {\tt PYTHIA} generator in this region as
compared to analytical calculations \cite{ZFITTER}, partly to
four-fermion processes which were not taken into account completely, and partly
to reconstruction problems in real events with jets in the forward direction. 
If such excess events
in the data pass the later steps of the selection, 
the background is likely to be
underestimated and the limits derived in the absence of a signal
are thus conservative. In the final
data sample the background from Z($\gamma$) events is rather unimportant,
however.  

In the continued selection events were rejected if there was a 
neutral particle, either with an energy
above 60~\GeV, or isolated from the nearest jet by at least 20\dgree, and with
an energy above 20~\GeV. These criteria served to remove radiative return events.

To reduce the WW background, events were rejected if they had a charged particle 
with momentum greater than 20~\GeVc\ or if the most isolated electron or muon (if any) 
had momentum greater than 10~\GeVc\ or was more than 20\dgree\ from the nearest jet.
Figure \ref{fig:DATAMC1}(c) shows the distributions of
transverse momentum (\pT) divided by $\sqrt{s}$
for real and
simulated data, after the above selection.

In the last step of the selection, events were accepted if they satisfied
any of the following three sets of criteria, optimised for different neutralino
mass differences (\DM). 
The criteria involved the transverse momentum (\pT),
longitudinal momentum (\pL), and invariant mass (\Mvis) of the visible
system, as well as the mass recoiling against it (\Mrec).
Also the 
acollinearity of the two jets and their
scaled acoplanarity 
were used in this step. The events were accepted if:

\begin{itemize}
\item[(i)] 
      $\Mvis < 0.1\sqrt{s}/c^2$, $\Mrec > 0.7\sqrt{s}/c^2$, and
      $\pT > 7~\GeVc$. In addition, the
      scaled acoplanarity was required to exceed 40\dgree. 
      These criteria are efficient for low \DM\ ($\sim$10~\GeVcc). 

\item[(ii)] 
      $0.1\sqrt{s}/c^2 < \Mvis < 0.3\sqrt{s}/c^2$, $\Mrec > 0.6\sqrt{s}/c^2$, and
      $\pT > 8~\GeVc$. The scaled acoplanarity had to exceed 25\dgree.
      These criteria are efficient for intermediate \DM\ ($\sim$40~\GeVcc).

\item[(iii)] 
      $0.3\sqrt{s}/c^2 <\Mvis < 0.5\sqrt{s}/c^2$, $\Mrec > 0.45\sqrt{s}/c^2$,
      $12~\GeVc <\pT < 35~\GeVc$, and $\pL < 35~\GeVc$.
      The scaled acoplanarity had to exceed 25\dgree, and the acollinearity had
      to be below 55\dgree.
      These criteria are efficient for high \DM\ ($\sim$90~\GeVcc).

\end{itemize}

Figure  \ref{fig:DATAMC1} (e) shows a comparison of the scaled
acoplanarity for the real and simulated data events passing
the last step of the selection.

\subsection{Acoplanar leptons search\label{sub:2L}}

The search for acoplanar leptons selects events with exactly two
isolated oppositely charged particles (lepton candidates) with momentum above 1~\GeVc, and
at most five charged particles in total. 

This search was slightly modified with respect to
Ref.~\cite{charneut183}, as follows. The minimum number of TPC pad rows
required for the two selected charged particles was increased from four to five.
The lepton identification requirements were changed,
accepting as electrons those particles which had
an associated energy in the electromagnetic calorimeter exceeding half of
the measured momentum, while for muons the ``loose tag'' 
criteria \cite{DELPHI} were used. Either, both particles in the pair were 
required to be selected as electrons and
not simultaneously identified as muons, or else both particles had to
be muons. In addition to the acoplanarity, the acollinearity between the
two particles also had to exceed 10\dgree. The minimum transverse momentum
required
was increased from 5 to 6~\GeVc, and the maximum accepted energy in the
STIC was reduced from 1 to 0.3~GeV. To improve the rejection of WW background,
events with missing momentum above 45~\GeVc, and a scalar sum of the
momenta of the two selected particles in excess of 100~\GeVc, were rejected. 
Prior to the last step of the selection 65 real data events were accepted, 
while
the expected background was 62.8$\pm$4.4 events, with a contribution of
31.1$\pm$0.6 events from \WW\ production.
As in Ref.~\cite{charneut183}, the last step involved 
three sets of criteria sensitive 
to different \DM\ ranges. These criteria were 
unchanged, except for the
minimal missing mass required in the selection optimised for large \DM, 
which was changed from 0.4$\sqrt{s}/c^2$ to 0.2$\sqrt{s}/c^2$.

Figure  \ref{fig:DATAMC1} (b,d,f) shows a comparison between
real and simulated data for events passing the initial step of the 
above selection corresponding to rejection
of Bhabha events (b), passing the intermediate step corresponding
to rejection of two-photon events (d), and passing the last step (f).
Real and simulated data were in good agreement throughout.

\subsection{Multijet search\label{sub:MJET}}

The multijet search was optimised for cascade decays of neutralinos with large
mass splittings, giving high energy jets. Events with energetic
photons, characteristic of the decay \XN{2}~$\to$~\XN{1}$\gamma$,
were subjected to less stringent selection criteria, giving a separate
set of selected events with low background and comparatively high efficiency.

\begin{figure}[ht]
\begin{center}
\mbox{\epsfysize=18.0cm\epsffile{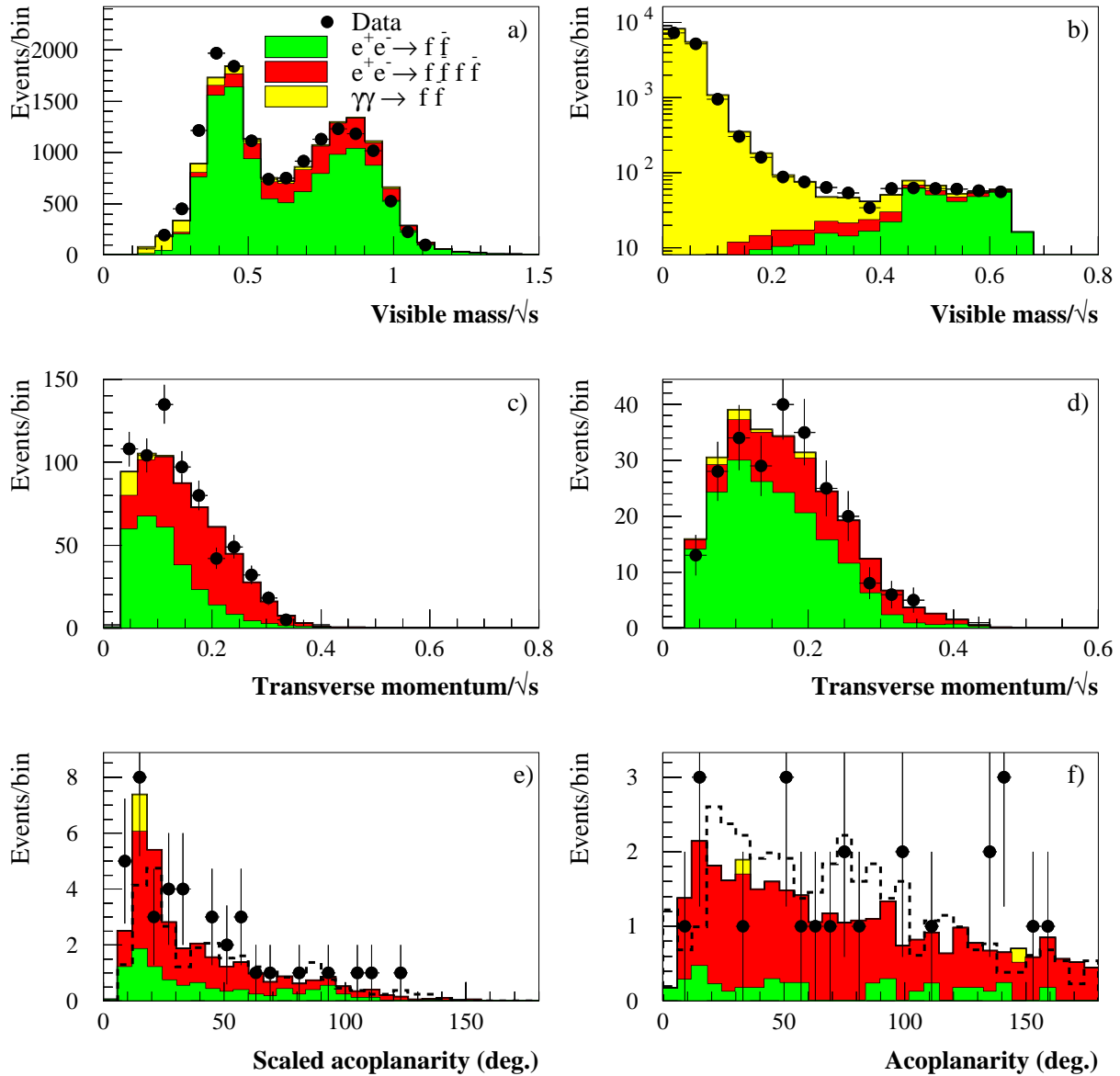}}
\caption[Data-MC comparison]{
The  comparison between the real and simulated data
for the multijet  selection (a,c,e) and multilepton  selection
(b,d,f) is shown at three different stages of the selection.
Plots (a,b) show the \Mvis\ divided
by the centre-of-mass energy at an initial stage of the
selections. Plots (c,d) show the missing transverse momentum
at an intermediate stage of the selections. Plots (e,f) show 
the acoplanarity after the last step of the selections.
The selections are described in sections \ref{sub:MJET} and
\ref{sub:MLEP}.
The distributions expected for \XN{2}\XN{3} production with 
\XN{3}$\to$\XN{2}\ffbar$\to$\XN{1}\fpfbarp, normalised to a cross-section
of 2~pb, are also shown for decays into quark and lepton pairs in
e) and f), respectively (dashed histograms). Equal mass differences 
$\MXN{3}\! -\! \MXN{2}\! =\! \MXN{2}\! -\! \MXN{1}\! =$~25~\GeVcc\ were
assumed.    
 }
\label{fig:DATAMC2}
\end{center}
\end{figure}

At least five well-reconstructed charged particles were required, and at least
one of these had to have a transverse momentum exceeding 2.5~\GeVc.
The transverse energy of the event had to be greater than 25~GeV, and
the total energy of tracks which were badly reconstructed or did not originate 
from the interaction point
was required to be less than 30~GeV and less than 45\% of the visible energy.
In addition, the calorimeter energy associated to such tracks had to be less
than 20\% of \Evis.
Figure  \ref{fig:DATAMC2} (a) shows the distributions of
\Mvis\ divided by the centre-of-mass
energy  for real data and simulated background events
passing the above selection.
The excess of ``radiative return'' events observed in the acoplanar jets 
search is visible also here, and the comments of section \ref{sub:2J}
apply. Similarly, the deficit of events in the real data
with $\Mvis/\sqs$ 
close to unity can be partly explained by a known excess 
of {\tt PYTHIA} events with little initial state radiation.

The total energy in the electromagnetic calorimeters had to be less than
70~GeV, and there had to be no single calorimeter shower above 60~GeV.
The energy carried by particles within 30\dgree\ of the beam had to be less
than 60\% of the visible energy.
The total visible
energy had to be less than 135~GeV, the polar angle of the total momentum had to satisfy
$|\cos\thetap| < 0.9$, and the transverse momentum had
to exceed 6~\GeVc.
Figure  \ref{fig:DATAMC2} (c) gives a comparison of the
\pT/$\sqrt{s}$-distributions for real data and simulated background
following the above selection.

The scaled acoplanarity (see section \ref{sub:2J}), calculated 
forcing the number of jets to two, had to be greater than 10\dgree.
The polar angle of
the most energetic jet had to be outside the range 
between 85\dgree\ and 95\dgree\ to avoid an insensitive detector 
region close to 90\dgree,
and its energy had to be less than 56~\GeV.

To reject WW background it was required that there be no charged particle with
a momentum above 30~\GeVc, and that the momentum of the most
isolated electron or muon (if any) be below 10~\GeVc, or below 4~\GeVc\ if the
angle between the lepton and the nearest jet was greater than 20\dgree.

Events with a photon signature were then selected on the basis of reconstructed
photons in the polar angle range between 20\dgree\ and 160\dgree, isolated by more
than 20\dgree\ from the nearest charged particle track. 
If there was only one such photon its energy was required to be between 10~GeV 
and 40~GeV; with more than one photon, at least two had to have energy greater 
than 10 GeV.

For the complementary sample, without a photon signature, two additional requirements
were imposed to reject \Zn$\gamma$ events: the mass recoiling against the system of
visible particles had to be greater
than 100~\GeVcc, and all jets with energy above 20~GeV had to have a ratio of energy
in charged particles to energy in neutral particles which was above 0.15.

Lastly, events selected by the searches for acoplanar jets or leptons
(sections \ref{sub:2J} and \ref{sub:2L}) were rejected.
Figure  \ref{fig:DATAMC2} (e) shows the acoplanarity distributions
for real and
simulated events without a photon signature passing the last step of the
selection.

\subsection{Multilepton search \label{sub:MLEP}}

The multilepton search is sensitive to cascade decays involving leptons,
which can dominate if there are light sleptons.

The first step in the selection, in common with
the tau cascade and low $E_T$ searches
(sections \ref{sub:XT} and \ref{sub:4T}), was as follows.  
The number of charged particles was required to be at least two and at most eight,
and events with more than four neutral particles were rejected.
The reconstructed invariant mass had to
be below 120~\GeVcc, and the recoil mass above 20~\GeVcc.
The calorimeter energy
associated to particles which were badly reconstructed or did not originate at
the vertex, $E_{\mathrm bc}$, was required not to exceed 0.4~\Evis, while the
energy of well reconstructed charged particles had to be
greater than 0.2~\Evis.
It was also required that $\Evis\! +\! E_{\mathrm bc}\! <\! 140~\GeV$.

In the following step, at least two charged particles were required to be
identified leptons.
Figure  \ref{fig:DATAMC2} (b) shows a comparison between \Mvis/$\sqrt{s}$
distributions for real and simulated events passing the above selection.

To reject \Zn$\gamma$, two-photon, and Bhabha events, the transverse 
momentum of the event was required to exceeded 8~\GeVc, and the polar
angle of the total momentum to satisfy $|\cos \theta_p| < 0.9$.
The transverse energy of the event had to be greater than 25~\GeV, and the energy
in the STIC was required to be less than 10~GeV.
The distributions of \pT/$\sqrt{s}$ for real and
simulated data, following the above selection, are compared in
figure ~\ref{fig:DATAMC2} (d).

For events with exactly two isolated well-reconstructed charged
particles the following requirements were imposed.
The acoplanarity and acollinearity of these two particles had to exceed
15\dgree\ and 6\dgree, respectively. If the total energy in
electromagnetic calorimeters exceeded 50~GeV the acollinearity was required to
be greater than 10\dgree. To reject W pairs decaying leptonically
it was required that the product of charge and cosine of polar angle was 
less than $-0.1$ for each of the two charged particles.

For events with two reconstructed jets, the scaled 
acoplanarity was required to be greater than 15\dgree.

Lastly, events selected by the searches for acoplanar jets or leptons
(sections \ref{sub:2J} and \ref{sub:2L}), or by the
multijet search (section \ref{sub:MJET}), were rejected.
Figure  \ref{fig:DATAMC2} (f) shows the distributions of acoplanarity
for real and simulated data, following the above selection.

\subsection{Tau cascade search \label{sub:XT}}

The tau cascade search is sensitive to \XN{1}\XN{2} production
with $\XN{2}\! \to\! \stau\tau$ and $\stau\! \to\! \XN{1}\tau$,
where the second $\tau$ produced has very low energy.
The first step of the selection was the same as for the multilepton
search (section \ref{sub:MLEP}),
with the additional
requirement of no more than two reconstructed jets.
Two or more of the charged particles also
had to satisfy stricter criteria on reconstruction and impact parameters.

\begin{figure}[ht]
\begin{center}
\mbox{\epsfysize=18.0cm\epsffile{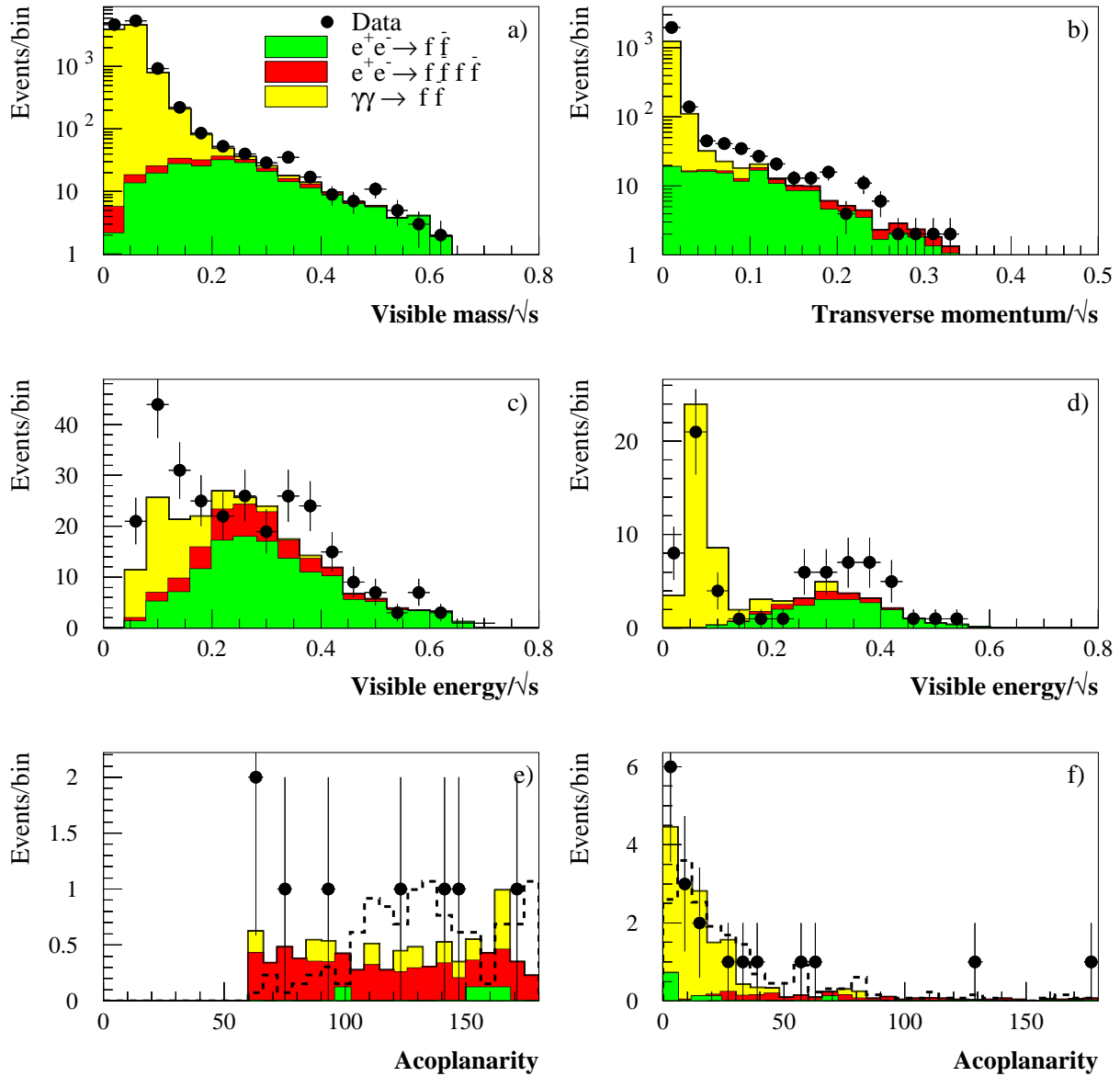}}
\caption[Data-MC comparison]{
The  comparison between the real and simulated data
for the tau cascade selection (a,c,e) and low \ET\ selection
(b,d,f) is shown. Plots (a,b) show the \Mvis\ divided
by the centre-of-mass energy and transverse momentum
divided by the centre-of-mass energy
at an initial stage of the
selections.
Plots (c,d) show the visible energy divided
by the centre-of-mass energy
at an intermediate stage of the selections.
Plots (e,f) show the acoplanarity after the last step of the selections.
The selections are described in sections \ref{sub:XT} and
\ref{sub:4T}.
The dashed line in e) shows the tau cascade signal 
expected from \XN{1}\XN{2} production with 
$\XN{2} \to \tau \stau \to \tau \tau \XN{1}$ ,
\MXN{1}~=~34.8~\GeVcc, \mstau~=~36.8~\GeVcc, and \MXN{2}~=~60~\GeVcc.
In f) the dashed line corresponds to \XN{2}\XN{2}
production with \XN{2}$\to$\XN{1}\ellell ($\ell$~=~e,$\mu$,$\tau$),
\MXN{1}~=~35~\GeVcc, and \MXN{2}~=~40~\GeVcc. The signals are normalised to 
2~pb. 
 }
\label{fig:DATAMC3}
\end{center}
\end{figure}

In the next step, the highest and second highest momenta of charged particles
were required to be below
50~\GeVc\ and 25~\GeVc, respectively, and at least one charged particle had to
have a transverse momentum above 2.5~\GeVc.
Events with neutral showers above 300~MeV
within 20\dgree\ of the beam axis were rejected.
The visible mass distributions, for real and
simulated data at this stage of the selection,
are compared in figure \ref{fig:DATAMC3} (a).

The criteria to reject \Zn$\gamma$, two-photon, and Bhabha events, were
the same as for the multilepton search (section \ref{sub:MLEP}), except for
the minimum transverse momentum which was reduced to 7~\GeVc, and the removal
of the transverse energy requirement.

Figure \ref{fig:DATAMC3} (c) shows distributions of \Evis/$\sqrt{s}$
as a comparison between real and
simulated data, selected with the above criteria.
There is an evident excess in the energy region dominated by two-photon 
interactions. This has been studied in a recent workshop on 
generators at LEP2 \cite{LEPGENWS}. The background from
two-photon interactions giving hadronic final states is known to 
be underestimated, and the
process \gamgam $\to\! \ellell$ is also not well described by simulation.
In the case of \tautau\ the treatment of tau decays in the generator
was approximate, and polarisation effects were absent. Furthermore,
some four-fermion processes were not completely accounted
for in the simulation.
If the two-photon background the end of the selection is also underestimated
the obtained limits are conservative, but in any
case this background is not the dominant one.

Events with exactly two isolated, well-reconstructed, oppositely
charged particles were required to have acollinearity and acoplanarity
above 60\dgree. The smaller of the two momenta had to be below 70\% of
the greater one, and below 10~\GeVc.

For events with two reconstructed jets the scaled acoplanarity 
(see section \ref{sub:2J})
was required to be greater than 20\dgree, and the acoplanarity and
the acollinearity greater than 60\dgree.

Lastly, events selected by the searches for acoplanar leptons or jets
(sections \ref{sub:2L} and \ref{sub:2J}) or the
multilepton search (section \ref{sub:MLEP}) were rejected.
Figure  \ref{fig:DATAMC3} (e) shows the
acoplanarity distribution for events passing the
complete selection, in real data and simulated background.

\subsection{Low transverse energy search \label{sub:4T}}

The low transverse energy (\ET) search was designed to complement 
the multilepton
search for cascade decays or \XN{2}\XN{2} production
with low mass splitting where \XN{2}~$\to$~\XN{1}\ellell. The first step of
the selection was the same as for the multilepton search.
In the second step, it was required that there be at least three and at
most five charged
particles, and that all had momenta above 500~\MeVc.
Two or
more of the charged particles
had to satisfy stricter criteria on reconstruction and impact parameters.

In the third step, the highest and second highest momenta of charged particles
were required to be below
50 and 25~\GeVc, respectively. At least one charged particle had to
have a transverse momentum above 2.5~\GeVc, and at least one had to
be an identified lepton. There had to be no neutral shower
within 20\dgree\ of the beam axis, and the second highest jet
energy had to be below 30~GeV.

Figure  \ref{fig:DATAMC3} (b) shows the distributions of \pT/$\sqrt{s}$
for events fulfilling the above criteria in the real and simulated
data. Excess data events from two-photon interactions and the 
``radiative return'' process are visible here too. Again, this could 
give too conservative limits if such excess events were to survive
the complete selection. The overall effect of the low transverse 
energy search on the obtained limits is rather small, however.

The bulk of the two-photon background was rejected by the requirements 
that the polar angle of the total momentum had to 
satisfy $|\cos \theta_p| < 0.9$, and that the transverse energy of the 
event had to be greater than 4~\GeV. The distributions of \Mvis/$\sqrt{s}$
for the real and simulated data, following these requirements, are 
compared in figure \ref{fig:DATAMC3} (d).

The specific requirements for events with exactly two well
reconstructed, isolated, charged particles were the same as in
section \ref{sub:MLEP}, with the
additional requirement
that at least one of the tracks had to have a
momentum below 15~\GeVc.

Events with transverse momentum exceeding 8~\GeVc\ and
transverse energy greater than 10~\GeV\ were rejected, unless
the scaled acoplanarity, calculated forcing the number of
jets to two, was above 20\dgree.

Lastly, events selected by the searches for acoplanar jets or leptons
(sections \ref{sub:2J} and \ref{sub:2L}), the
multilepton search (section \ref{sub:MLEP}),
or the tau cascade search (section \ref{sub:XT}) were rejected.
Figure  \ref{fig:DATAMC3} (f) shows the distributions of scaled
acoplanarity
for real and simulated events passing the complete selection.

\section{Selected events and expected backgrounds \label{sec:EVENTS}}

Table \ref{tab:EVENTNUM} shows the number of events selected in
the different searches in real data and the numbers expected from
the Standard Model background. Also shown are the main background sources
contributing in each channel and
the typical efficiency of each search for MSSM points where it
is relevant. 

\begin{table}[ht]
\begin{tabular}{|l|r|r|c|c|}
\hline
Search & Data & Total bkg. & Main bkg. & Typical eff. (\%)\\ \hline
Acoplanar jets           & 19  &  21.0$\pm$1.6  &
                         \WW,ZZ  &  10 -- 30   \\
Acoplanar electrons        & 16  &  20.7$\pm$3.7  &
                         \WW,\gamgam  &  10 -- 40   \\
Acoplanar muons           & 16  &  14.6$\pm$1.3  &
                         \WW,\gamgam  &  10 -- 40   \\
Multijets, $\gamma$:s   &  2  &   4.3$\pm$0.5  &
                         \Zn$\gamma$  &  10 -- 20  \\
Multijets, no $\gamma$:s  & 39  &  31.8$\pm$1.9  &
                         \Zn$\gamma$, \WW &  10 -- 40 \\
Multileptons             & 23  &  28.2$\pm$1.2  &
                         \WW  &  30 -- 50   \\
Tau cascades             &  8  &   9.0$\pm$1.0  &
                         \WW,\gamgam($\to \mu^+\mu^-$) &  13 -- 19  \\
Low \ET                & 18  &  19.0$\pm$3.3  &
                         \gamgam($\to \tau^+\tau^-$)  &  7 -- 10   \\ \hline
\end{tabular}
\caption{\small
Results of the different searches.
The typical efficiency of each
search for MSSM points where it is relevant is shown. The efficiencies
depend typically on the masses of the sparticles involved
in the process.
For any given search, events are
explicitly rejected if accepted by one of the searches appearing
earlier in the table.
}
\label{tab:EVENTNUM}
\end{table}

The main reason for
the variation of the efficiencies is the variation
of the masses of the particles involved in the process.
The explicit rejection
of events to avoid overlapping selections
limits the efficiencies
for those searches in which such
rejection is performed (see section \ref{sec:SELECT} and table
\ref{tab:EVENTNUM}).
The total number of events selected in the different searches was 141, 
with $149\! \pm\! 6$ background events expected.
The errors given for the background estimates are due to the finite
sizes of the simulated background samples. No error was assigned to
account for the excesses of data events seen at early stages of the
selections. 
In conclusion, the results are in good agreement with the
expectation from Standard Model background, and no indication of a 
signal was found.

\section{Signal efficiencies and upper limits \label{sec:RESULTS}}
In the absence of a signal, cross-section limits were derived 
based on the efficiencies for simulated neutralino events.
A total of 360\,000 \XN{1}\XN{2} events
was simulated
for 108 different combinations of masses
with \MXN{1} and \MXN{2} ranging from 5~\GeVcc\ to 90~\GeVcc\ and
from 20~\GeVcc\ to 180~\GeVcc, respectively, and for
different \XN{2} decay modes (\qqbar\XN{1}, \mumu\XN{1}, \ee\XN{1},
\stau$\tau$ ).
A further 100\,000 \XN{2}\XNN{3}{4}\  events with cascade decays,
were simulated for 56 different points. In addition, about 5$\cdot$10$^8$
events were simulated using {\tt SGV} in order to obtain signal efficiencies
for about 10$^5$ MSSM points. 

\begin{figure}[ht]
\begin{center}
\mbox{\epsfysize=18.0cm\epsffile{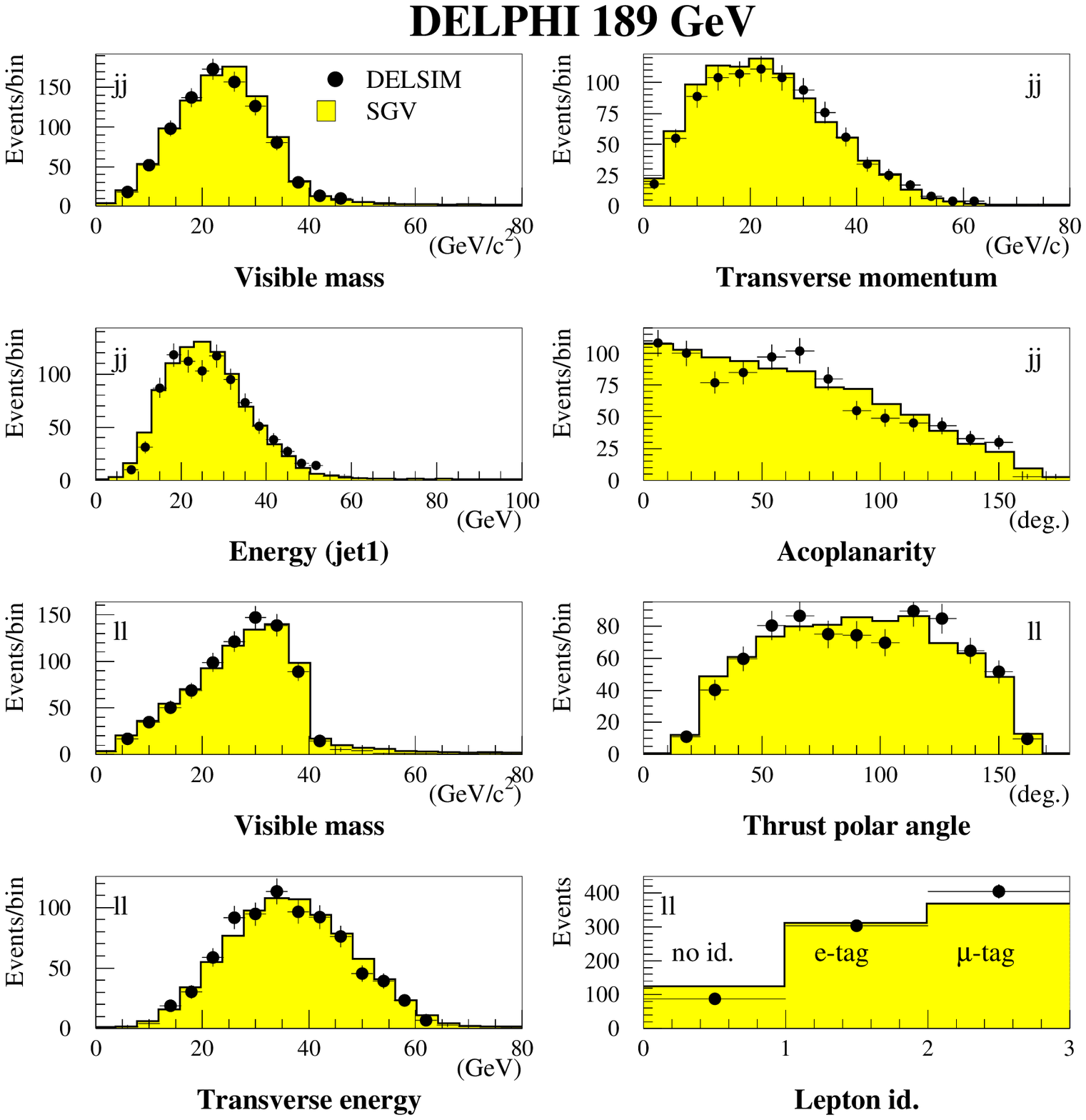}}
\caption[sgv]{
The expected distributions of relevant event variables
characterizing  the  \XN{1}\XN{2} production with $\MXN{1}= 40$ \GeVcc\
and $\MXN{2}= 80$ \GeVcc,
as obtained using the full detector simulation ({\tt DELSIM}) and {\tt SGV} for
the acoplanar jet (jj) topology (upper four plots) and acoplanar lepton (ll) 
topology (lower four plots).
The decays $\XN{2} \to \XN{1}  q \bar{q}$ or \XN{1}\ellell\ were assumed
as appropriate (\ellell\ denotes \ee\ and \mumu\ in equal proportions).
}
\label{fig:sgv1}
\end{center}
\end{figure}

\begin{figure}[ht]
\begin{center}
\mbox{\epsfysize=18.0cm\epsffile{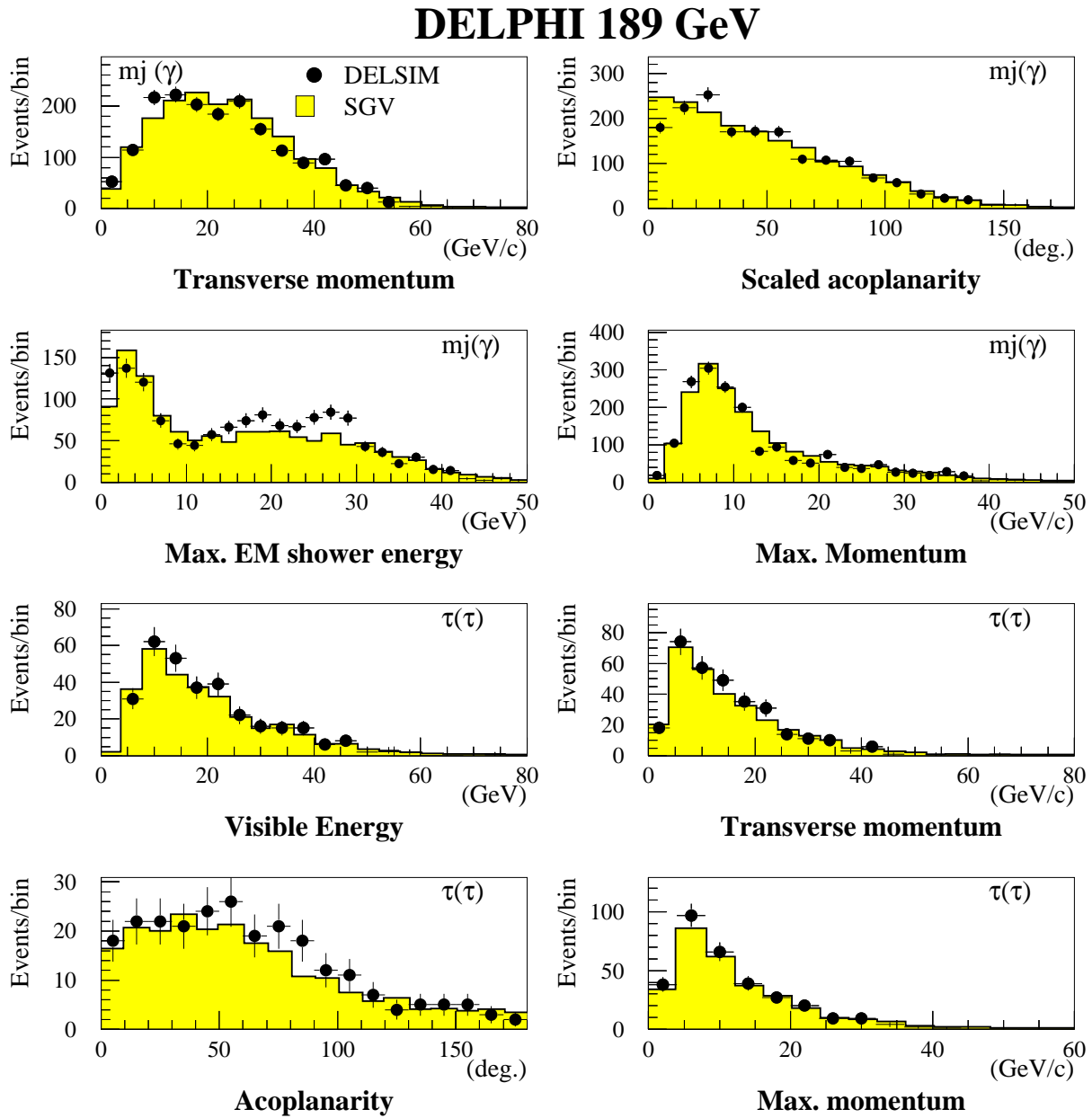}}
\caption[sgv]{
The expected distributions of relevant event variables
characterizing neutralino production for the multijet topologies
(upper four plots) and the tau cascade topology (lower four plots). 
In the multijet case chosen,
\XN{4}\XN{2} production dominates with 50\% of the \XN{2} decaying
to \XN{1}$\gamma$. The neutralino masses are \MXN{1}~=~31~\GeVcc,
\MXN{2}~=~60~\GeVcc, and \MXN{4}~=~100~\GeVcc. In the tau cascade 
case, \XN{1}\XN{2} production with 
$\XN{2} \to \tau \stau \to \tau \tau \XN{1}$ was assumed with
\MXN{1}~=~34.8~\GeVcc, \mstau~=~36.8~\GeVcc, and \MXN{2}~=~60~\GeVcc.
 }
\label{fig:sgv2}
\end{center}
\end{figure}

\begin{figure}[ht]
\begin{center}
\mbox{\epsfysize=18cm\epsffile{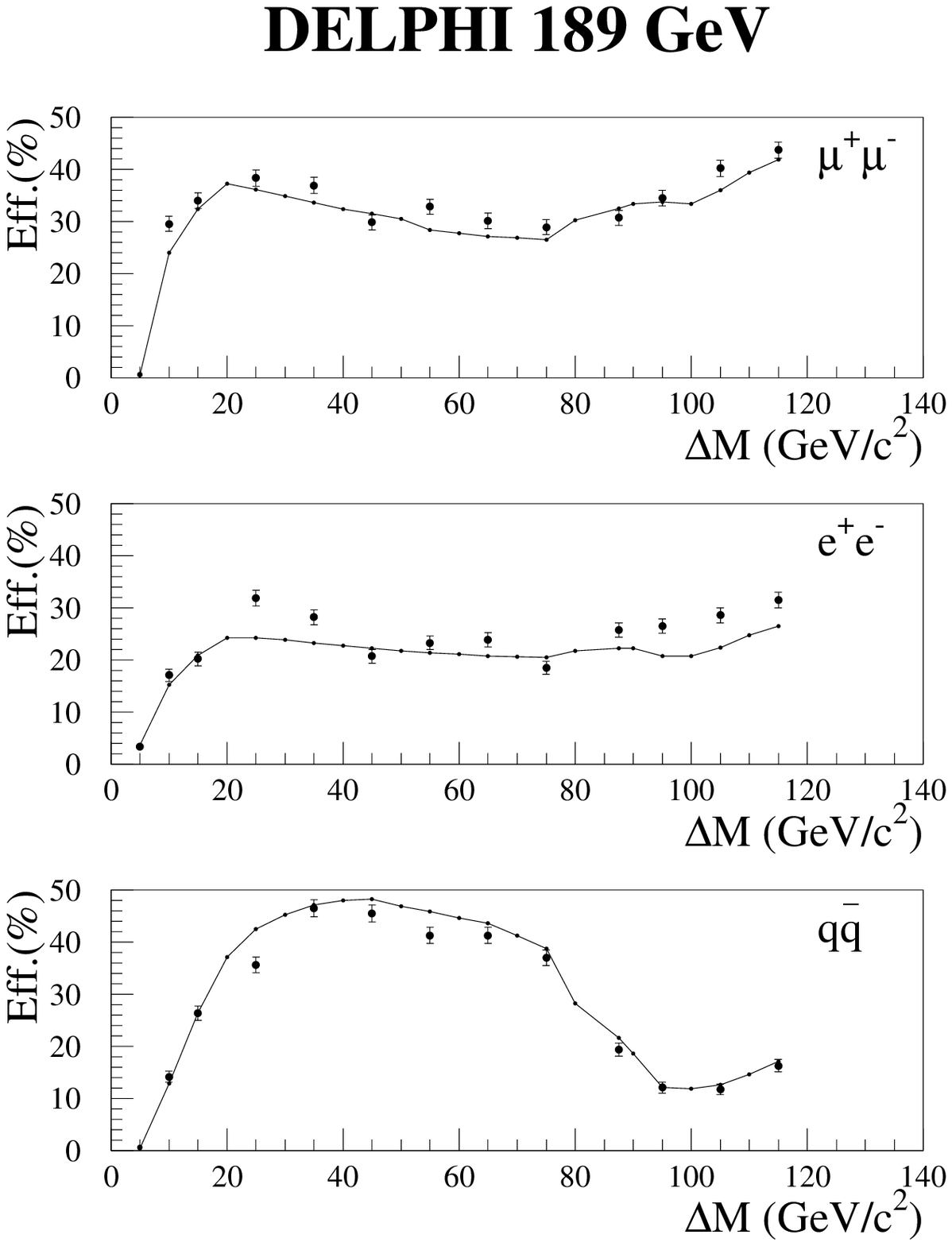}}
\caption[MSSM limits in ($\mu$,$M_2$) plane]{
Efficiencies for \XN{1}\XN{i} production as 
obtained with the full simulation ({\tt DELSIM}, points with
error bars)
and the fast simulation ({\tt SGV}, points connected by
straight lines) for 
different $\DM\! =\! \MXN{i}\! -\! \MXN{1}$, assuming the decays
\XN{i}$ \to $\XN{1}\ffbar\ (f=$\mu$,e,q). The mass of
\XN{1} was fixed to 35~\GeVcc.
}
\label{fig:DMSGV}
\end{center}
\end{figure}

Figures \ref{fig:sgv1} and \ref{fig:sgv2}
show the expected distributions for some relevant event variables
for \XN{1}\XN{2} production
as obtained using the full detector simulation and {\tt SGV}.
The efficiencies obtained using SGV agreed typically to $\pm$10\% 
relative with 
those obtained by full simulation. 
Figure \ref{fig:DMSGV} shows a
comparison between {\tt SGV} efficiencies (curves)
and those from the full simulation (points) as a function of \DM\ 
in the topologies with acoplanar leptons and acoplanar jets. In
the case of leptonic events the SGV efficiencies are generally 
lower, giving conservative limits. In the hadronic case the SGV
efficiencies tend to be higher, and they were therefore 
conservatively reduced by 20\% in the limit calculations. The effect 
on the \MXN{1} limit for $\tanb\! \!= 1$ \cite{LSPLIM} was found to 
be completely negligible. 

\begin{figure}[ht]
\begin{center}
\vskip 1cm
\begin{center}
\mbox{\epsfysize=7.5cm\epsffile{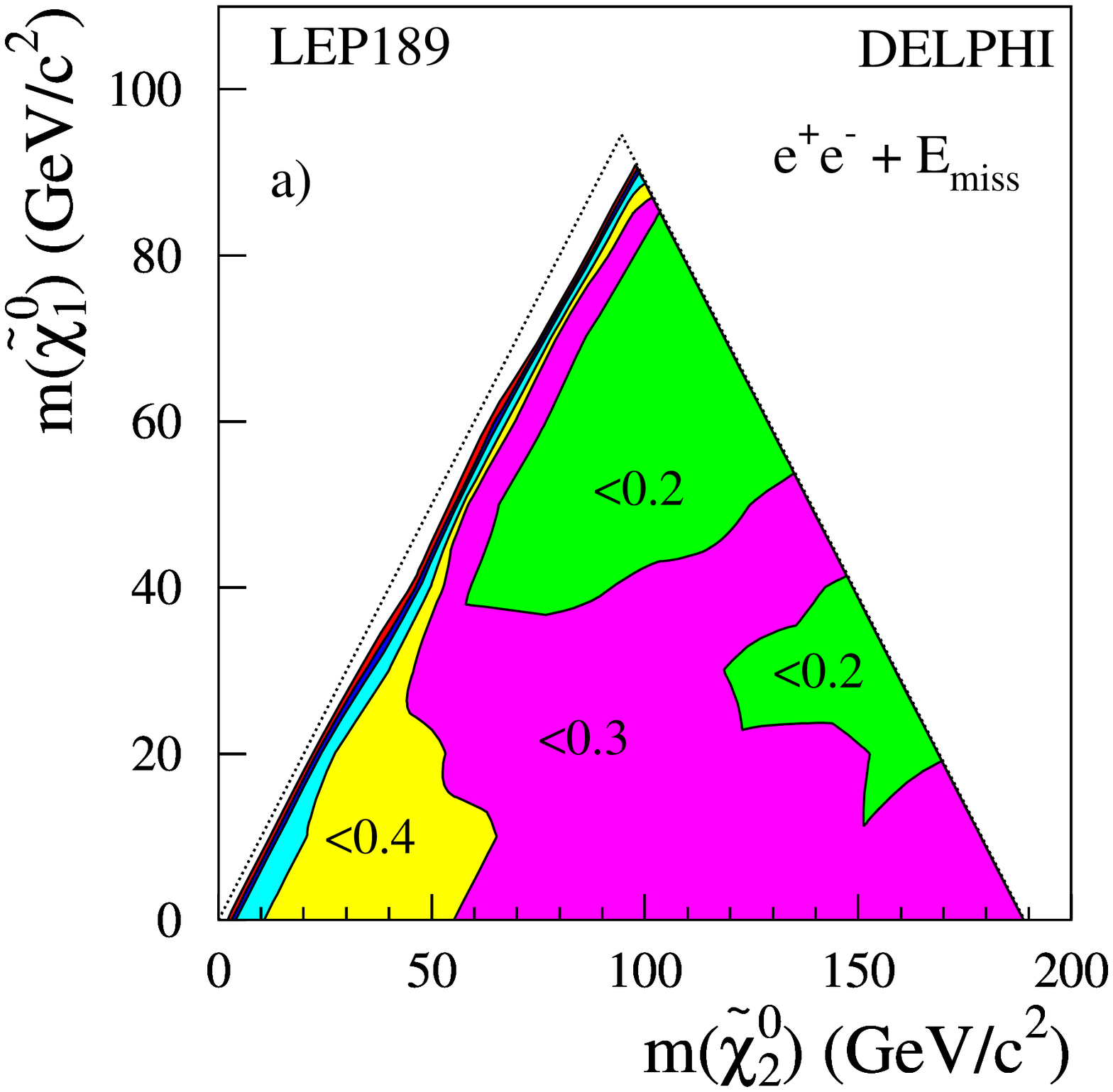}}
\mbox{\epsfysize=7.5cm\epsffile{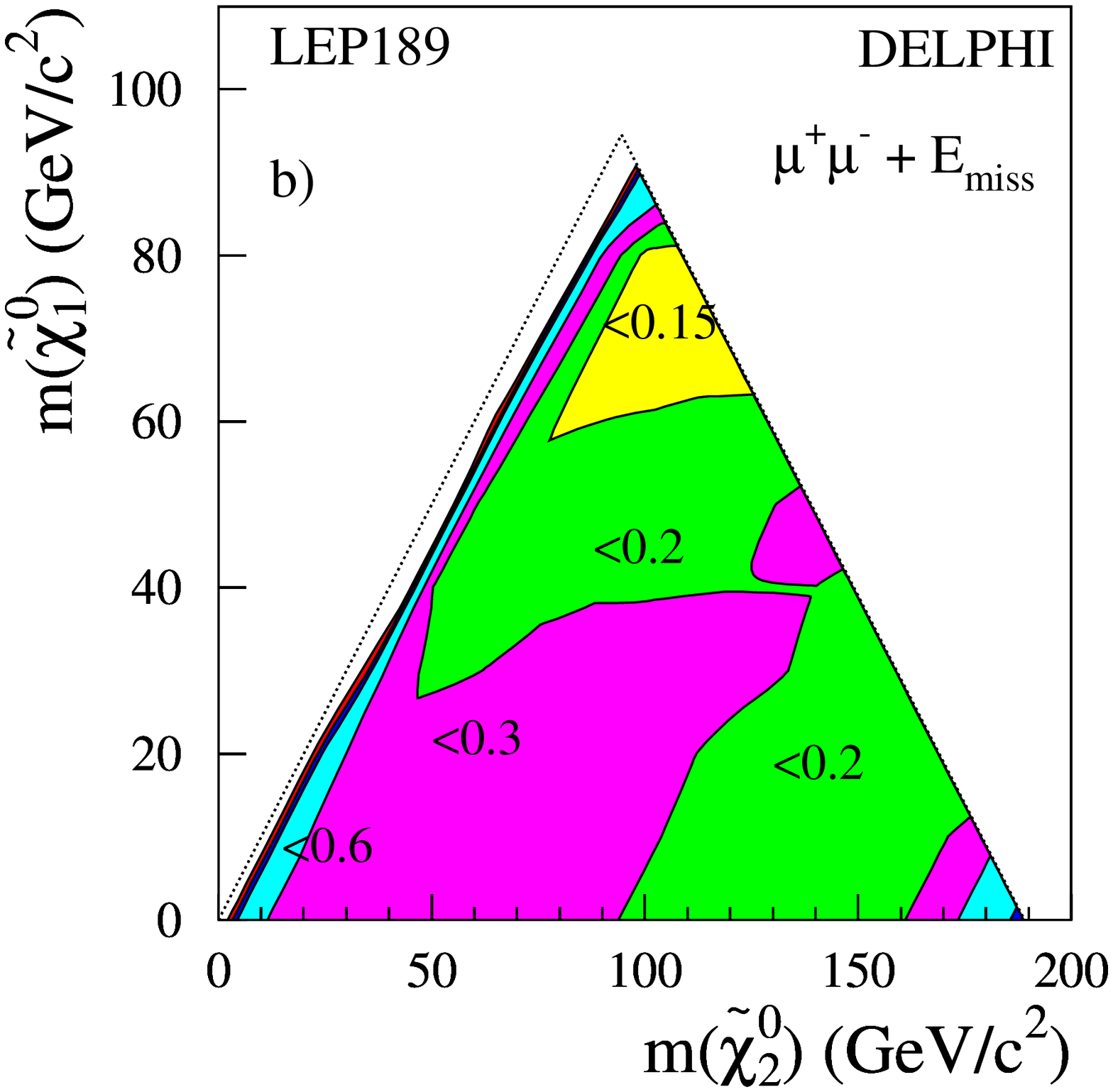}}
\mbox{\epsfysize=7.5cm\epsffile{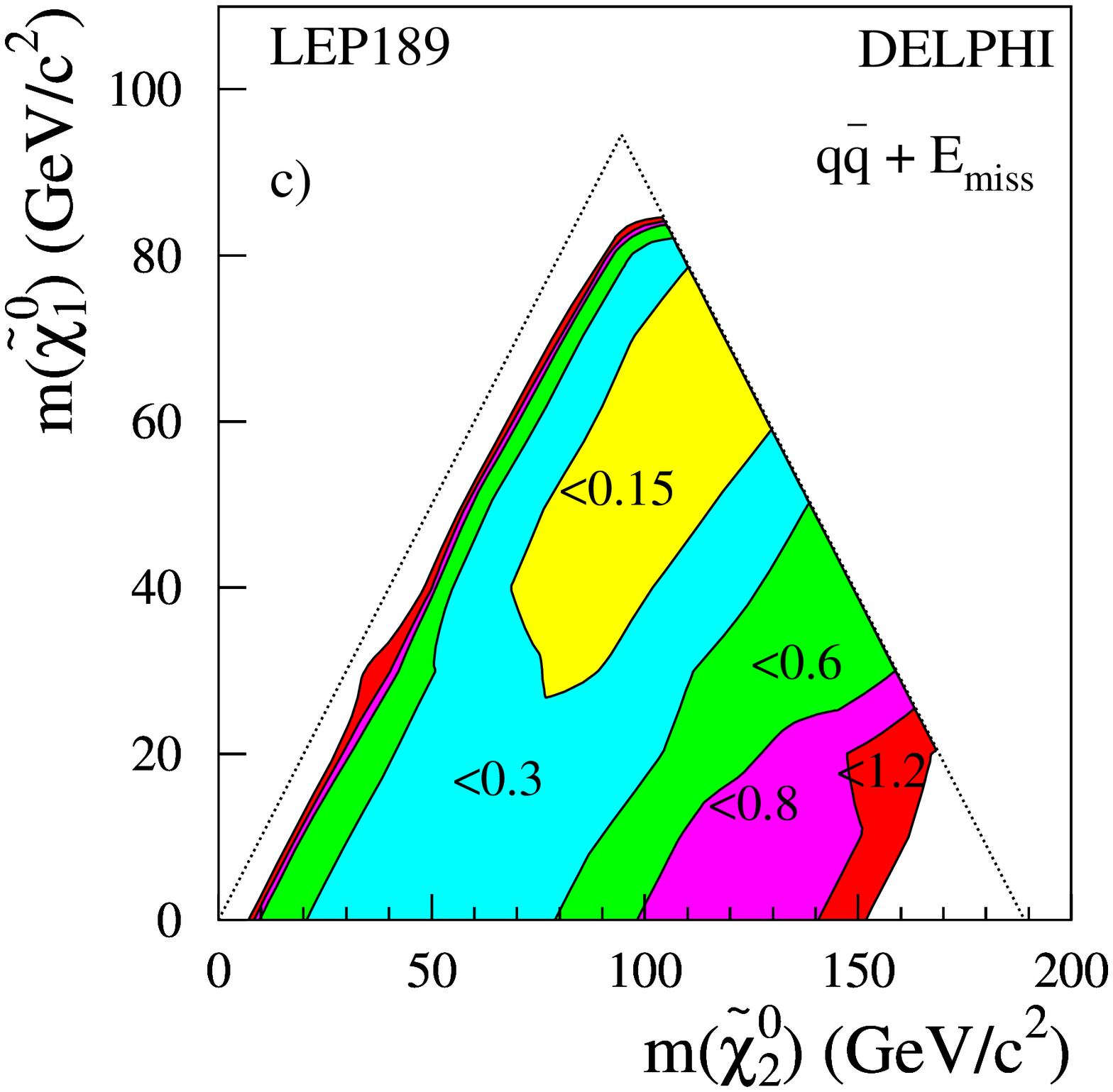}}
\mbox{\epsfysize=7.5cm\epsffile{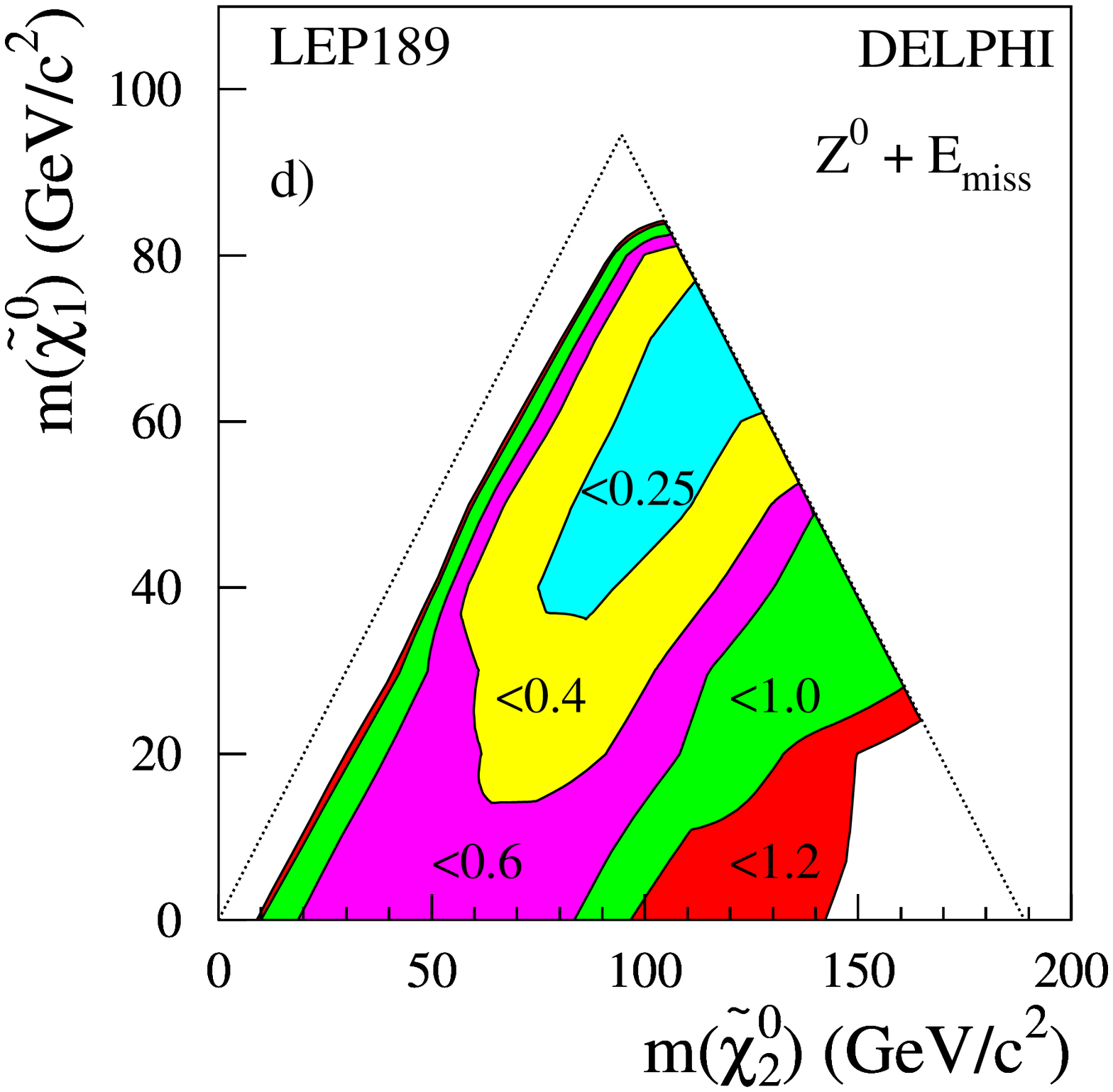}}
\end{center}
\vskip 1cm

\caption[Model independent cross-sections]{
Contour plots of upper limits on the cross-sections
at the 95\%
confidence level for \XN{1}\XN{2} production at $\rs~=~189~\GeV$. In each plot,
the different shadings correspond to regions where the cross-section limit in
picobarns is below the indicated number.
For figures a), b), c), \XN{2} decays into \XN{1} and a) \ee,
b) $\mu^+\mu^-$, and c) \qqbar, while in d) the branching
ratios of the \Zn ~was assumed, including invisible states.
The dotted lines indicate the kinematic limit and the
defining relation $\MXN{2}>\MXN{1}$.
}
\label{fig:M1M2LIM}
\end{center}
\end{figure}

\begin{figure}[ht]
\begin{center}
\mbox{\epsfysize=11.0cm\epsffile{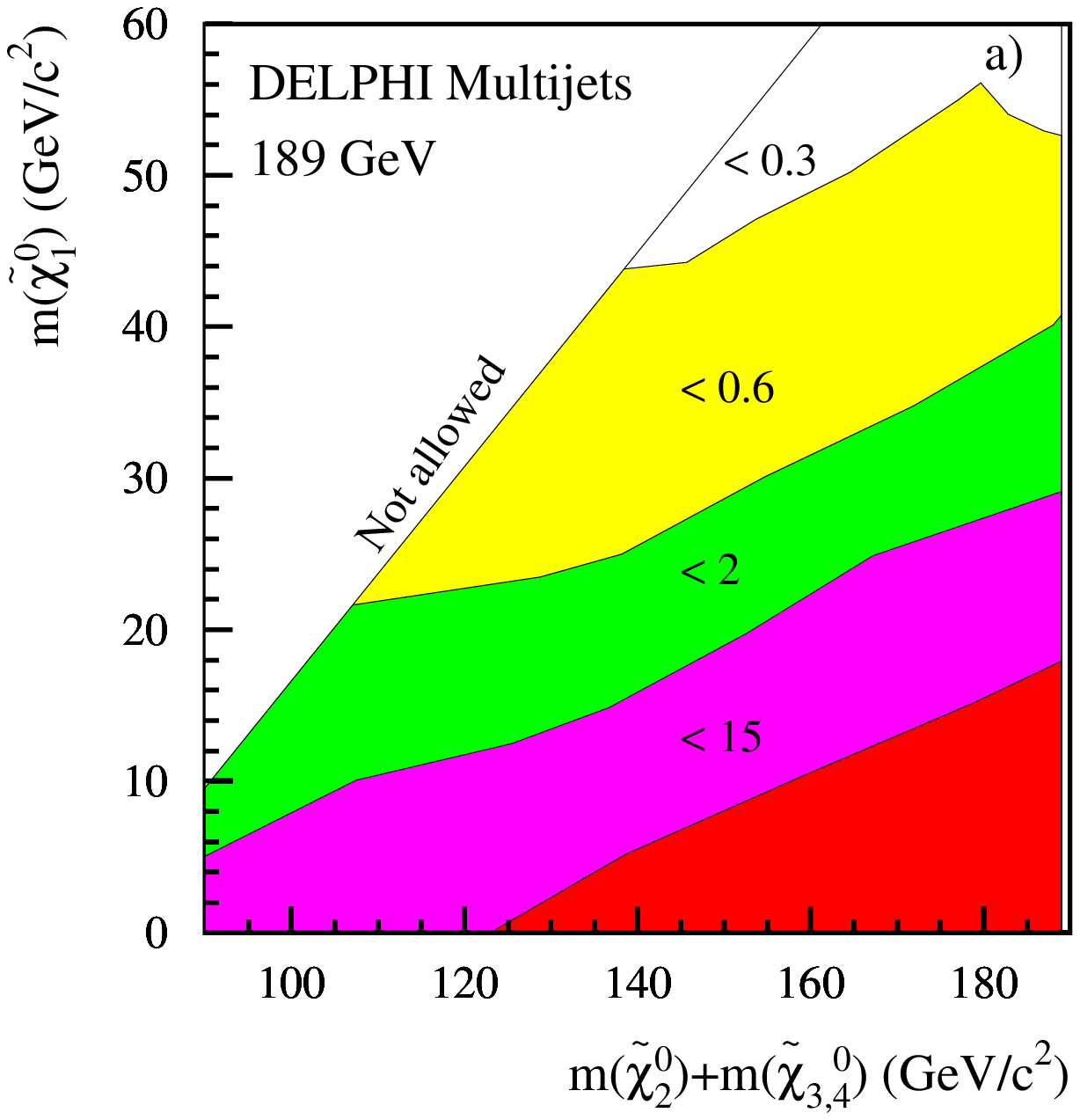}}
\vskip -1.0cm
\mbox{\epsfysize=11.0cm\epsffile{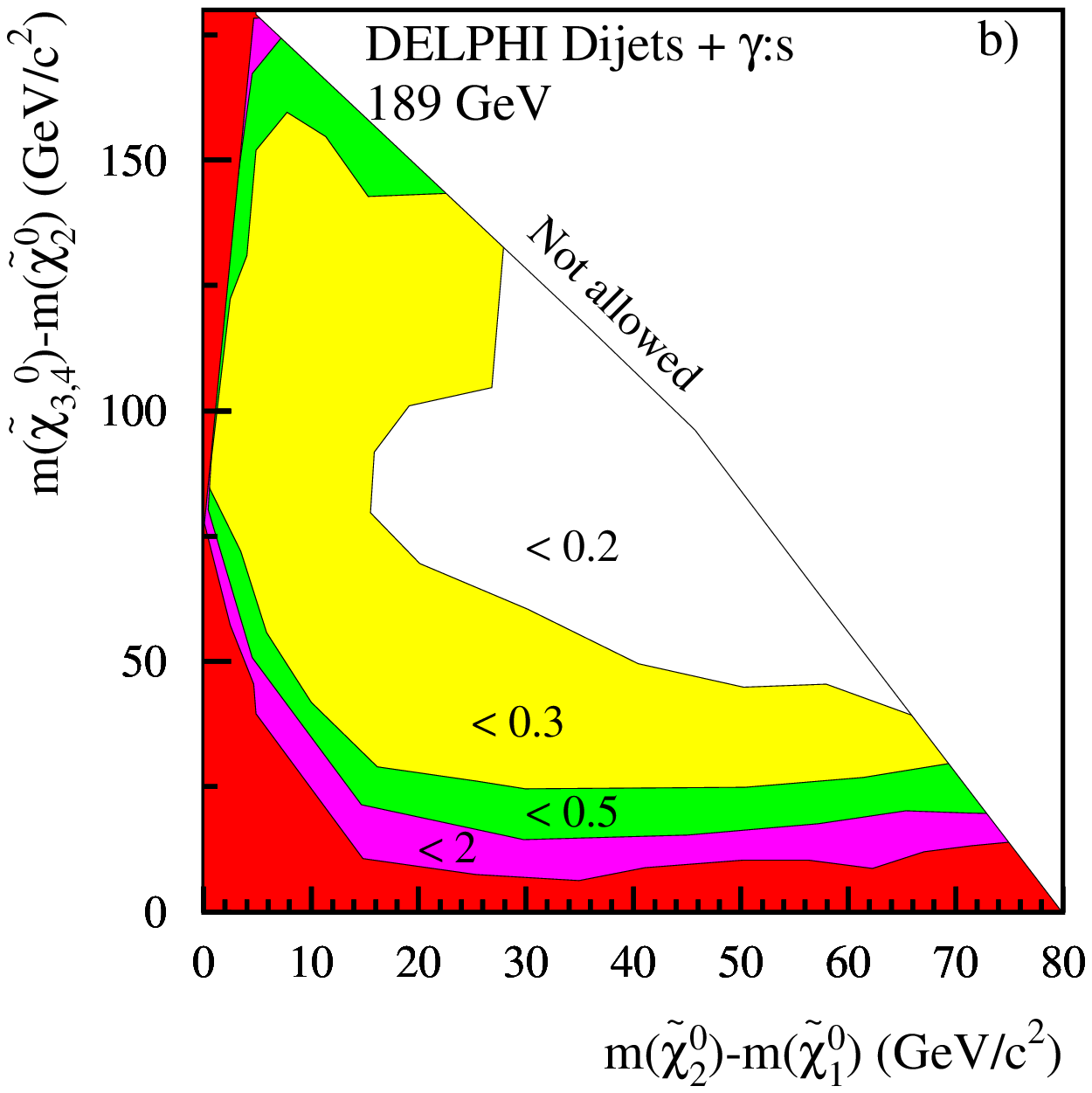}}
\vskip -0.25cm
\caption[Multijet cross-sections]{
Upper limits on the cross-sections
at the 95\%
confidence level for \XN{2}\XN{i} production with 
\XN{i}$\to$\XN{2}\qqbar\ ($i$=3,4) at $\rs\!~=\!~189\!~\GeV$. 
The different 
shades correspond to regions where the cross-section limit in
picobarns is below the indicated number. In the darkest shaded
regions there are points which are not excluded for any 
cross-section.
\XN{2} was assumed to
decay into \XN{1}\qqbar\ in a), and into \XN{1}$\gamma$ in b).
The limits in a) are based on the acoplanar jets and multijets
selections, while those in b) derive from the search for
multijets with photons.
}
\label{fig:MJXS}
\end{center}
\end{figure}

The limits for the \XN{1}\XN{2} production,
as obtained from the searches for acoplanar leptons and jets, are
shown in Figs.~\ref{fig:M1M2LIM} assuming different branching ratios.
Similarly, Figs.~\ref{fig:MJXS}(a,b) show cross-section
limits for \XN{2}\XN{i} production ($i\! =$ 3 or 4). 
For each mass combination, the limits
were obtained by examining many possible ($\mu$,$M_2$) points
for several \tanb\ values and high $m_0$,
where \XN{2}\XN{i} production was kinematically allowed. The
point giving the worst limit was taken. In the white
regions marked ``Not allowed'', no such points were found. Figure~\ref{fig:MJXS}(a)
shows the limit obtained using a Bayesian combination \cite{OBRA} of the results
from the multijet and acoplanar jet searches in the case where
\XN{i}$\to$\XN{2}\qqbar\ and \XN{2}$\to$\XN{1}\qqbar.
Figure~\ref{fig:MJXS}(b) gives the corresponding
limits when \XN{2}$\to$\XN{1}$\gamma$,
as obtained from the search for
multijet events with a photon signature.

\begin{figure}[ht]
\begin{center}
\mbox{\epsfysize=18cm\epsffile{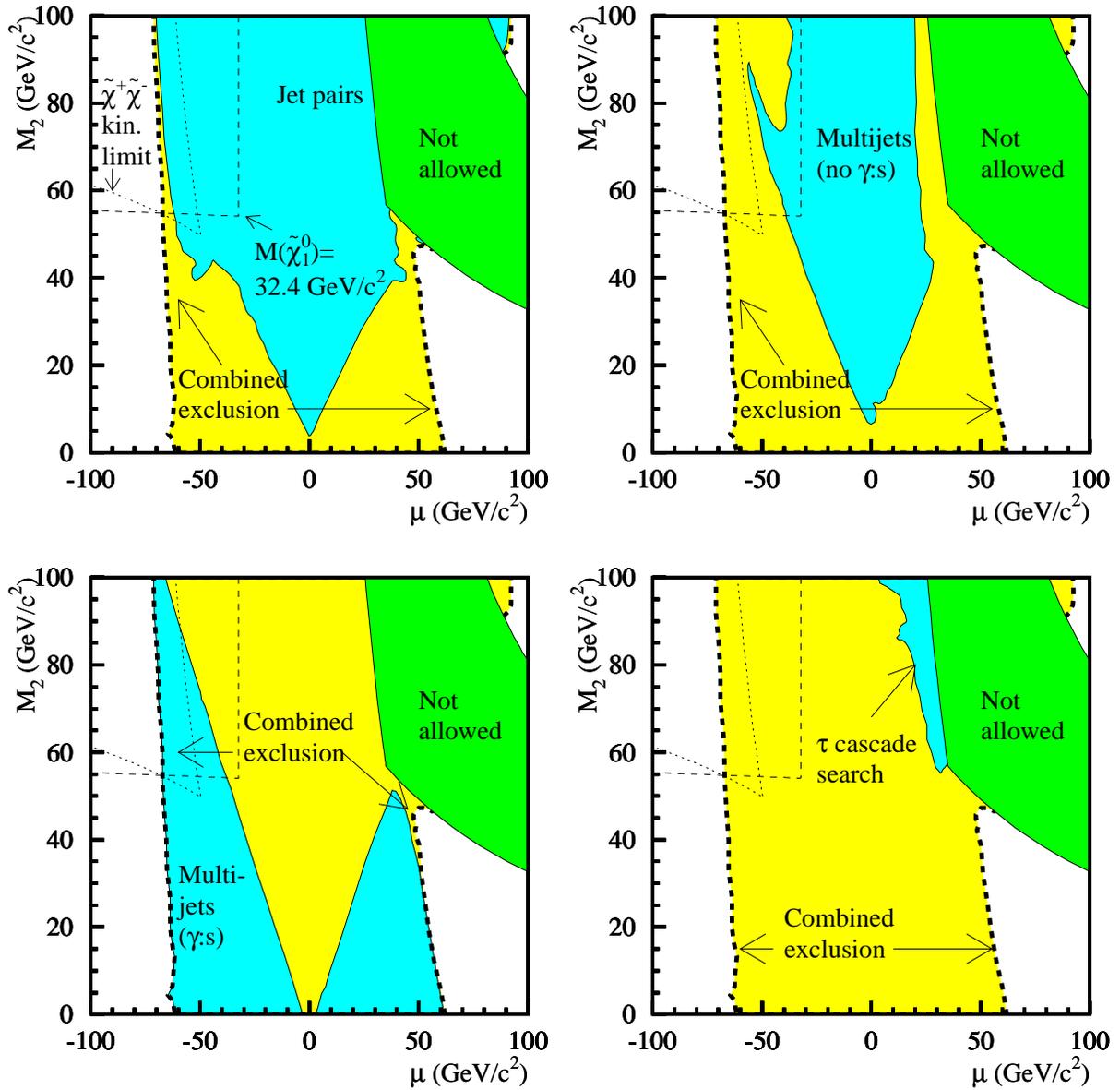}}
\caption[MSSM limits in ($\mu$,$M_2$) plane]{
Regions in the ($\mu,M_2$) plane excluded at 95\% confidence level
for \tanb =1, assuming $m_0$~=~1~\TeVcc. The exclusion by 
individual searches for jet pairs (top left), 
multijets without $\gamma$:s (top right),
multijets with $\gamma$:s (bottom left), 
and $\tau$ cascades (bottom right) are compared with 
the combined exclusion based on all searches
(thick dashed curve and light shading). Also shown are the 
kinematic limit for chargino production (thin dotted curve) and the 
isomass contour for the minimum allowed neutralino mass \cite{LSPLIM}
(thin dashed curve). In the region marked ``Not allowed'' the
chargino is the LSP.
}
\label{fig:M01K}
\end{center}
\end{figure}

\begin{figure}[ht]
\begin{center}
\mbox{\epsfysize=18cm\epsffile{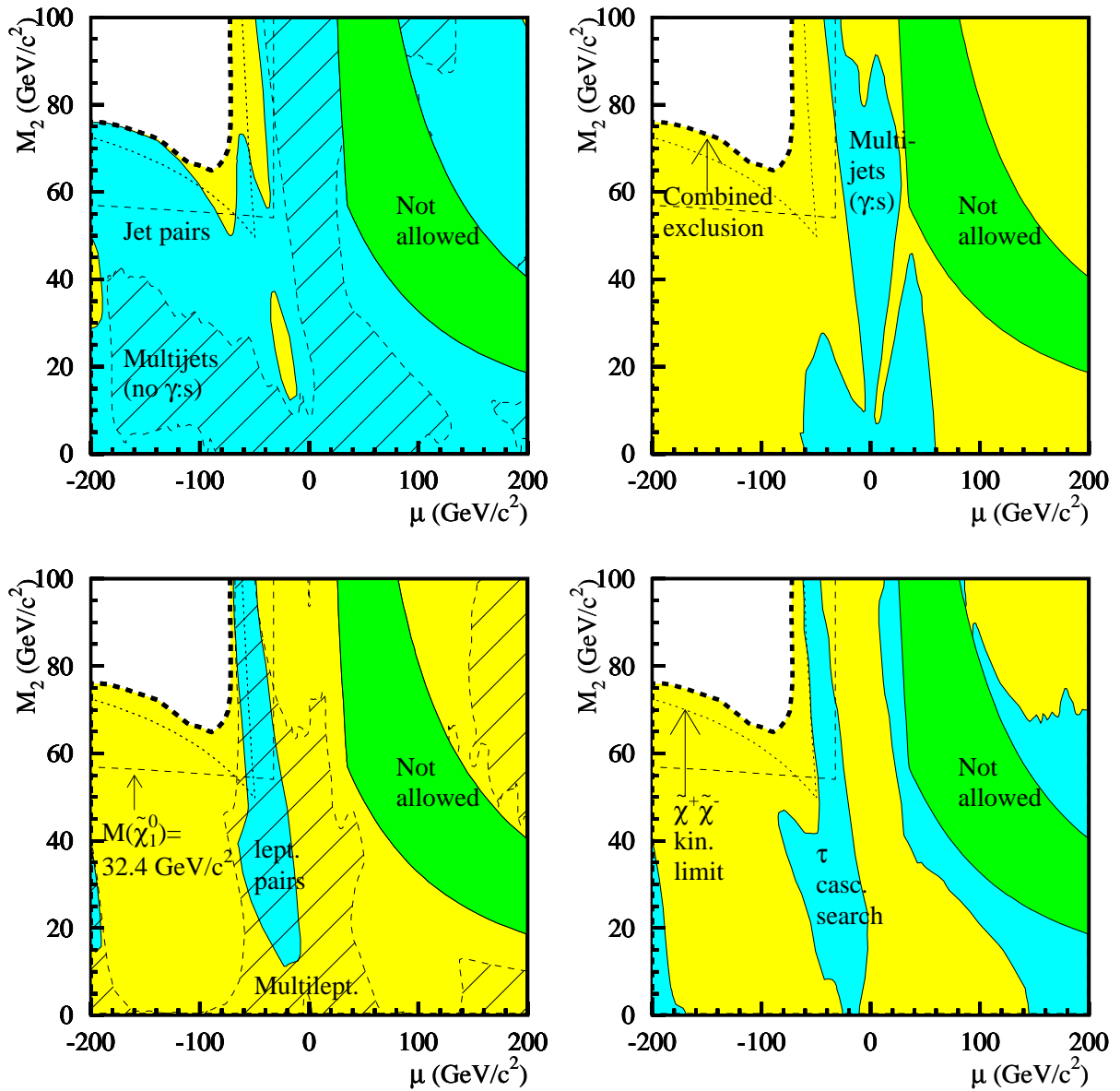}}
\caption[MSSM limits in ($\mu$,$M_2$) plane]{
As figure \ref{fig:M01K}, but for $m_0$~=~80~\GeVcc\ and
six different contributing searches. From left to right
and top to bottom: jet pairs and multijets
without $\gamma$:s (hatched), multijets with $\gamma$:s, lepton
pairs and multileptons (hatched), and $\tau$ cascades. 
}
\label{fig:M080}
\end{center}
\end{figure}

In addition to such limits on the production cross-sections, the approach
using a fast simulation makes it possible to scan regions of the MSSM
parameter space and calculate the efficiencies
directly at each point, simulating all neutralino
production channels and decay chains. Since they were defined to be
mutually exclusive, the different selections can be combined
using the Bayesian multi-channel approach \cite{OBRA} to obtain the exclusion
confidence level for each set of MSSM parameters\footnote{The same
procedure is applied in Ref.~\cite{LSPLIM}, including also the production
of other supersymmetric particles.}. 
Figs.~\ref{fig:M01K} and \ref{fig:M080} show the regions excluded by the
different contributing searches in the ($\mu$,$M_2$) plane for \tanb~=~1 and
$m_0\!~=\!~1\!~\TeVcc$ and 80~\GeVcc, respectively. Also shown are the
combined exclusion regions for the two values of $m_0$.
In the region indicated as ``Not allowed'' the
lightest chargino is lighter than \XN{1}. 
Although the process for which it was designed is not important here,
the $\tau$ cascade search is efficient for cascade decays involving 
leptons in the region close 
to $\mu\!~=\!~0$ for $m_0\!~=\!~80\!~\GeVcc$,
and when the chargino-neutralino mass difference is small.
(In the latter case the decay $\XNN{2}{3}\! \to \XPM{1}\ell\nu$ is followed by
an almost invisible chargino decay.)
 
The thin dotted curve in the figures indicates the chargino isomass
contour corresponding to the kinematic limit for \XP{1}\XM{1} production. 
For high $m_0$ this is very close to the exclusion limit from chargino
searches. For low $m_0$ the region excluded from chargino production
is smaller \cite{LSPLIM}, but the neutralino excluded region is 
increased, as can be seen from Figure~\ref{fig:M080}.
Therefore the overall
limit on \MXN{1} for \tanb~=~1 is determined by the intersection of
the chargino isomass contour with the region excluded by neutralinos for
high $m_0$ \cite{LSPLIM}. The corresponding \XN{1} isomass contour is
shown as the thin dashed curve.

At low $m_0$ and low $M_2$,
the region excluded by neutralinos shrinks with increasing \tanb\
due to  enhancement of the invisible $\XN{2} \to \snu \nu$, $\snu \to \nu \XN{1}$
decay channel. There is no substantial change of the high $m_0$ exclusion
region with the increase of \tanb.

\section{Summary \label{sec:SUMMARY}}

Searches for neutralinos at \rs~=~188.7~GeV, using several mutually
exclusive sets of criteria, gave no indications of a signal.
As a consequence,
upper limits on cross-sections for different
topologies were derived, ranging from about 0.1~pb to several picobarn.
The efficiencies computed with a full simulation of the DELPHI detector
were extended to the whole range of the SUSY parameters explored by
using a fast detector simulation, which included all neutralino production
and decay channels.
Exclusion
regions in the MSSM parameter space were
then derived.
The methods used were designed for deriving general MSSM mass
limits in the Minimal Supersymmetric Standard Model, as done in
a separate
letter \cite{LSPLIM}.

\subsection*{Acknowledgements}
\vskip 3 mm
 We are greatly indebted to our technical 
collaborators, to the members of the CERN-SL Division for the excellent 
performance of the LEP collider, and to the funding agencies for their
support in building and operating the DELPHI detector.
We acknowledge in particular the support of the 
Austrian Federal Ministry of Science and Traffics, GZ 616.364/2-III/2a/98, 
FNRS--FWO, Belgium,  
FINEP, CNPq, CAPES, FUJB and FAPERJ, Brazil, 
Czech Ministry of Industry and Trade, GA CR 202/96/0450 and GA AVCR A1010521,
Danish Natural Research Council, 
Commission of the European Communities (DG XII), 
Direction des Sciences de la Mati$\grave{\mbox{\rm e}}$re, CEA, France, 
Bundesministerium f$\ddot{\mbox{\rm u}}$r Bildung, Wissenschaft, Forschung 
und Technologie, Germany,
General Secretariat for Research and Technology, Greece, 
National Science Foundation (NWO) and Foundation for Research on Matter (FOM),
The Netherlands, 
Norwegian Research Council,  
State Committee for Scientific Research, Poland, 2P03B06015, 2P03B03311 and
SPUB/P03/178/98, 
JNICT--Junta Nacional de Investiga\c{c}\~{a}o Cient\'{\i}fica 
e Tecnol$\acute{\mbox{\rm o}}$gica, Portugal, 
Vedecka grantova agentura MS SR, Slovakia, Nr. 95/5195/134, 
Ministry of Science and Technology of the Republic of Slovenia, 
CICYT, Spain, AEN96--1661 and AEN96-1681,  
The Swedish Natural Science Research Council,      
Particle Physics and Astronomy Research Council, UK, 
Department of Energy, USA, DE--FG02--94ER40817.

\newpage

\newpage

\newpage

\newpage

\newpage

\newpage

\newpage

\newpage

\newpage

\newpage

\end{document}